# Dipolar many-body complexes and their interactions in stacked 2D hetero-bilayers

Xueqian Sun[1], Ermin Malic[2] and Yuerui Lu[1,3*]

[1]School of Engineering, College of Engineering and Computer Science, the Australian National University, Canberra, ACT, 2601, Australia

[2]Department of Physics, Philipps-Universität Marburg, 35037 Marburg, Germany

[3]Australian Research Council Centre of Excellence for Quantum Computation and Communication Technology, the Australian National University, Canberra, ACT, 2601 Australia

\* To whom correspondence should be addressed: Yuerui Lu (yuerui.lu@anu.edu.au)

**Abstract:**

**Highly customizable interfaces created by van der Waals stacked 2D materials provide an extremely flexible opportunity for engineering and effectively controlling material properties. The atomic-thin nature and strong scalability of transition metal dichalcogenides (TMDs), the star family of two-dimensional semiconducting materials, allow for the modulation of their inherent optical and electrical characteristics by utilizing various environmental stimuli. In such a material system, the stacking mechanism with spatial separation in the structure enables recent observations of dipolar many-body complexes with the interplay of multi-particles, leading to some exotic and novel excitonic phenomena and enabling the closer study of high-correlated quantum**

**physics. The presence of powerful dipole-dipole interactions among long-lived interlayer excitons can cause the system to enter unique classical and quantum phases with multi-particle correlations, such as dipolar liquids, dipolar crystals and superfluids. The strong binding energy of interlayer excitons in TMD-based hetero-bilayers especially enhances the critical temperature of these exotic phenomena. Here, we provide a concise summary of the recent frontier research progress on dipolar complexes and many-body effects in TMD double layers, encompassing fundamental theory and properties modulation. We reveal the significance and current challenges of this research field and present the potential developing directions of the hetero-bilayers in quantum physics and quantum devices by adding new levels of external control or integration.**

1. Introduction

Dipolar many-body systems, refer to collections of quasiparticles or entities with permanent electric dipole moments that interact through dipole-dipole interactions. The term "dipolar" arises from the presence of vertical electric dipoles associated with these particles, where positive and negative charges are separated by a finite distance, resulting in a net electric dipole moment[1-3]. Such systems exhibit inherent anisotropic many-body effects[4] arising from the interparticle interaction anisotropy within their structures. Dipolar excitonic states, particularly interlayer excitons (IXs)[5-10], have recently attracted significant attention and become a subject of study in the field of quantum technologies[11-13] and many-body physics[14-17]. The spatial separation between bound electron-hole pairs endows IX with a long lifetime, lasting in the microsecond range, owing to the significantly reduced oscillator strength stemming from the reduced overlap of the electron and hole wave functions, which decreases the radiative decay[18-21]. This provides sufficient time for them to efficiently undergo thermalisation after excitation, leading to strong exciton correlations and revealing a diverse range of fundamental collective many-particle effects governed by equilibrium quantum statistics[22], such as Bose-Einstein condensation (BEC)[23,24].

The theoretical prediction of the formation of exciton condensates in semiconductors dates back over five decades[25], with subsequent anticipation of a diverse range of classical and quantum phases[26-29] in this specific material system at cryogenic temperatures. In particular, the superfluid liquid of the excitonic system was initially proposed in a conventional semiconductor heterostructure with double quantum wells in GaAs[29], and progress in the fabrication of double quantum wells offered a favorable path for the experimental realisation of superfluidity in this system during the early 1990s. Thus, the study of the IX in coupled quantum wells initially gained attention as it provides a promising platform for exploring those quantum phenomena. However, despite the earlier demonstration of dipolar many-body systems in solid-state physics such as coupled-quantum well structures[22,30-34], the low binding energy of IXs in these conventional semiconducting structures limits their stability to ultra-low temperatures, typically in the range of only a few K[35,36]. This limitation poses a significant challenge and some classical phases theoretically demonstrated are still awaiting further validation through experimental results.

Interlayer excitons in 2D TMD heterobilayers exhibit a variety of similar fascinating characteristics that are not present in intralayer excitons. They possess the ability to overcome challenges faced in conventional semiconductors (Table 1), with binding energies more than one order of magnitude higher[20,37], surpassing the later improved binding energy of tens meV from GaN- and ZnO-based nanostructures[38,39], and inherit the advantages posed by graphene double layers for realising exciton condensate[40]. Owing to their strong binding energy and low mass of IX compared to classical commonly known bosons such as atomic gases[41-45], the degenerate Bose gas of IXs in TMD heterobilayers is expected to exhibit a macroscopically occupied quantum state at high temperatures, achieving a coherent many-body quantum state. They are expected to facilitate new remarkable discoveries due to the strong dipolar interactions and additional controllable twist angles, spanning a broad spectrum that includes the most prominent features of exotic moiré super-lattices[46-48], high-temperature condensations[23,24,41] and high-temperature superfluidity[41,49-51]. Additionally, the presence of an electric dipole moment and the emergence of moiré superlattice allow for effective tunability under applied external fields[52-57] and alignment adjustments. The ability to control dipolar excitonic states and their interactions presents exciting possibilities for exploring novel phenomena and intriguing correlated many-body states[58-60].

Dipolar coupling between dipolar excitons plays a crucial role in engineering their interactions[61]. This is different from contact-like or Coulomb interactions, which are highly anisotropic and long-range, and can shift from being repulsive to attractive based on the relative position and orientation of the dipoles, leading to the appearance of various exciton states. Similar to dipolar excitons, other many-body complexes such as interlayer trions[56,62] and interlayer biexcitons[63,64] produced from the excitonic effect based on van der Waals heterostructures, are essential for studying quantum physics and optoelectronic applications of 2D semiconductors, because they could bring the exciton system into a multi-correlation regime owing to the remarkable interactions between them, resulting in many-body effects and strong multi-particle correlations. They also provide possibilities for controlled nonlinear optics and spintronics applications. The formation of these complexes and the related collective behaviours, as well as phase transitions, were theoretically predicted, but experimental observations have only recently begun to emerge.

## 1.1 Overview

Although some recent reviews of purely dipolar excitons in physics can be found[37,65,66], there is a pressing need for a systematic and comprehensive review of dipolar complexes, encompassing other emergent many-body particles, to drive the development of correlated quantum phases and deepen our understanding of particle correlations and entanglement. Herein, we review the recent emergent excitonic phenomena in stacked 2D hetero-bilayers to explore quantum optics, with a particular focus on multiple many-body complexes and their dipolar interactions. We examine different exciton species separately in the following sections. The fundamental concepts and formation mechanisms of these various species will be initially introduced, followed by an exploration of their respective excitonic properties, dipolar interactions among these many-body complexes, and how these interactions influence their properties. Subsequently, the exotic quantum phenomena resulting from phase transitions will be elucidated. We present a contemporary understanding of various dipolar excitonic states and discuss the potential opportunities they offer by integrating excitonic physics.

## 2. Many-body complexes in 2D heterobilayers

Van der Waals (vdW) heterostructures can be assembled by vertically stacking different atomic thin layers of 2D materials[67-69], enabling coupling between two or multiple layers, to create the design of new materials with tailored and richer properties[18,70-73]. A comprehensive introduction to the various types of intrinsic many-body complexes exhibited in this structure is elaborated before delving into greater detail in the subsequent section.

### 2.1 Excitons

Strong bound excitons typically emerge in monolayer transition metal dichalcogenides (TMDs) under the trigger of light owing to strong Coulomb interactions. These are intralayer excitons (Fig. 1a), corresponding to the direct recombination transition of electrons and holes within single layers[74], with orders of magnitude shorter lifetimes than IX. These TMD excitons are tightly bound and stably preserved at room temperature, can strongly couple to light, and are endowed with many-body interactions[75,76]. However, when two TMD monolayers are brought together, a new interface can be directly formed with electrons and holes residing on opposite sides in different TMD layers, caused by type-II band alignment in hetero-bilayers[20,77-79] (Note S1). This results in spatially indirect interlayer excitons and fast transition processes[78,80,81].

### 2.1.1 Hybrid excitons

When intralayer and interlayer excitons interact with each other and become hybridized by sharing the same electron/hole due to the overlapped wavefunctions, a hybrid exciton is formed, exhibiting combined properties of intra- and interlayer states[82]. It typically emerges in homobilayers[17,53] but can also be formed in heterostrucures by properly choosing the TMD layers with nearly degenerate band edges, such as $MoSe_2/WS_2$[83] and $MoTe_2/MoSe_2$[84], thereby promoting the hybridization of intralayer and interlayer excitons. The degree of hybridization and concurrent many-body interactions can be effectively controlled through external fields[85].

Their charged counterparts and other complexes with similar vertical structure, such as interlayer trions and interlayer biexcitons, provide additional platforms for exploring quantum physics and adding dimensions to control exciton interactions.

### 2.2 Charged excitons

In this TMD type-II heterostructure, the lowest energy conduction band and the highest energy valence band are in different monolayers, with band offsets of approximately several hundred meV. Therefore, interlayer trions consist of two holes (electrons) in the same layer and one electron (hole) in the opposite layer for positively (negatively) charged interlayer trions because of the energetically favourable configuration[22,56], as depicted in Fig. 1b. Another type of interlayer trion with a dipole-free configuration was also predicted[86] and recently experimentally proven by a few reports[87,88], with negligible binding energy and formed owing to the sufficient free charges arising from substrates, defects, or gating.

### 2.3 Biexcitons

Interlayer biexcitons are generated by two bound interlayer excitons (Fig. 1c) with two parallel electric dipoles and the charges of the dipole are spatially separated by two different layers. The localized and free interlayer biexcitons feature opposite binding energies and are characterized by distinct interactions among individual dipolar excitons.

### 2.4 Moiré excitons

A distinctive capability of TMD heterobilayers to create a moiré superlattice enhances their prominence comparing with the traditional semiconductors. The moiré supercell can be introduced in stacked monolayer semiconductors by lattice mismatch and rotational misalignment between two layers[46,47,89-91]. It introduces an in-plane, nanoscale periodic potential landscape and is widely recognised as a useful mediator for manipulating the electronic band structure of materials[92-94], thereby significantly influencing the properties and dipolar interactions between IXs[95]. When the exciton Bohr radius is shorter than the moiré period constructed in two well-aligned monolayers, the IXs experience modulation under a spatially periodic potential and can move in the optical lattice[96], forming the new excitonic states known as moiré excitons. These excitons are influenced by the specific characteristics of the moiré pattern and exhibit unique optical and electronic properties[48,51,97-99].

### 2.5 Dipole-dipole interactions

All these complexes are endowed with a static, permanent out-of-plane electric dipole with the fixed orientation produced from the layer separation of their electrons and holes in real space, which results in long-range and strong dipole-dipole interactions[48,54,100] between the exciton states. However, distance between the dipoles (R), the separation between the centres, and the distance between different layers (*d*) can significantly affect exciton correlations. These distances of the intralayer excitons or in bulk semiconductors have been recognised as difficult to tailor, but they can become rather straightforward if the dipole is out-of-plane and consists of electron and hole from two layers[101].

Kis's group simply applied a 1L-hBN spacer between two TMD layers to enhance the size of the electrical dipole, demonstrating enhanced dipole-dipole interactions[57]. Fig. 1d reveals the dynamics of IXs, which are affected by partially counteracting mechanisms. Initially, because of the permanent dipole moments, the dipole-dipole interaction exhibits a strong repulsive inter-excitonic force, especially when the exciton density is increased, which is reflected by the density-dependent blueshift of the emission peak energy[28,102-104]. At a high density of excitons, changing the interlayer distance can effectively tune the dipolar interactions. This highlights the significance of controllable exciton formation efficiency (Box 1, Note S2). When the layer separation *d* is large, the dipole-dipole repulsion always purely dominates the dipolar interactions (Fig. 1e), and its strength varies with the distance. In contrast, when the distance between layers is decreased to a specific threshold value, the separation between dipoles (R) has a more significant impact on the manipulation of dipolar interactions among IXs, leading to a transition in the interactions from repulsive to attractive (black line in Fig. 1e, Note S3), which is predicted to be an exchange effect[28,105,106].

The interlayer exciton-exciton repulsive interactions are appreciably reduced with the emergence of attractive exchange interactions. This attractive dipole-dipole interaction, acting in the direction perpendicular to the dipole, should be distinguished from the attractive Coulomb interaction in the plane of excitons along the dipole axis. It not only promotes the appearance of other higher-order complexes, but also stimulates new phenomena such as pair superfluids[107] by generating interlayer dipolar bound pairs in 2D landscapes. Most progress has been reported in coupled quantum wells[30,34,108-111], and the attractive interaction has been recently experimentally evidenced[61]. However, TMD heterobilayers offer a promising alternative due to the additional degree of freedom they provide[21], allowing for properties engineering by controlling the layer separation and their relative stacking angle.

### 2.5.1 Interaction related binding energies of exciton complexes

The formation of interlayer biexcitons was predicted when modelling the coupled-quantum-well system as an ideal 2D layered structure[22,105]. Fig. 1f shows the calculated binding energies of the interlayer biexcitons with (green solid line) and without (blue solid line) dipolar interactions for comparison. The threshold interlayer separation is only found when considering exciton interaction[22], aligning well with *Meyertholen*'s stochastic variational method (SVM) calculations[112]. The critical distance ($d$) is reported to be 0.87 times the Bohr radius ($a_B$) of IX; thus, when $d$ is smaller than $0.87*a_B$, quantum-exchange-correlation effects occur, driving dipolar interactions to change from repulsive to attractive. This results in the formation of tightly bound biexciton phases formed by two IXs[22,41,105], as illustrated in Fig. 1f (green and red lines). The fabrication constraints of a coupled, double-quantum-well device posed challenges in achieving a small enough $d$ value. Conversely, atomically thin TMD heterobilayers can readily satisfy this condition. Calculations of trion binding energies (dark yellow line) revealed no critical distance due to the absence of a similar repulsive interaction. In addition to dipolar interactions, long-range Coulomb interactions provide another path for interlayer trions to interact with each other[56], which is also important for studying their dynamics. These multi-particles are crucial for creating an optically accessible platform for investigating many-body phenomena.

On the other hand, the physically fixed dipole distance ($d$) could change because the electron and hole wavefunctions may respond to dipolar interactions, leading to a distribution change and modification of the dipole moment. For example, the distance can be extended or shrunk when dipoles are approaching each other with the increasing repulsive interaction and the emergence of attractive interaction. However, the range of this variation should be very limited due to the strong confinement of electrons and holes, although the IX dipole could be tilted relative to the z direction[113]. Further changes are energetically unfavorable, and the wavefunctions spread for both electrons and holes in the z direction of the 2D structure is reported to be negligible[114].

The anisotropy and long-range correlations in dipolar excitonic systems yield novel quantum phases and topological states, facilitating exploration of multi-particle correlations and the interplay of different degrees of freedom in many-body systems. Strong binding energy in exciton complexes within TMD heterobilayers facilitates high-temperature condensation[23,24,41] and superfluidity[41,49-51]. Investigating this dipolar many-body system bridges the gap between condensed matter physics and quantum optics, providing valuable insights for both fields. Recent experimental advancements enable precise control of dipolar excitons, opening promising avenues for quantum computing[115], simulation[116], and sensing[117,118], given their relevance to quantum technologies[117,119] and information processing. As researchers continue to make progress in this field, systems can be engineered as quantum simulators, allowing exploration of more complex many-body quantum phenomena[120,121] that are challenging to investigate in natural systems.

## 3. Interaction-induced exciton properties
### 3.1 Dipolar interlayer exciton

Optical nonlinearity can be generated by mediating the interactions between photons, and the photon interactions at the single-quantum level are dominated by repulsion[54]. This interaction is significant for photon-based quantum devices and promotes the interaction of many-body states of light. Evidence of long-range interactions was recently experimentally reported with stable interlayer exciton dipoles[54], formed from the band alignment of two TMD single layers, $MoSe_2$ and $WSe_2$, with electrons in the $MoSe_2$ layer and holes in the $WSe_2$ layer. The emission energy of this oriented dipole can be modulated by the quantum-confined Stark effect and was tunable[122,123] with an external electric field (E) applied parallel to the dipole moment. As a signature of the dipolar repulsion, the energy position of the IX displayed an expected blue shift with the varying applied electrical field (Fig. 2a), whereas no shift was observed for a localised exciton peak from the $MoSe_2$ monolayer with the change in gate voltage because of the absence of an out-of-plane dipole moment for the intralayer exciton.

The blueshift feature in this dipolar exciton was corroborated by exciton density-dependent IX emission from both flat hetero-bilayer regions and those on a pillar[63]. In both cases, a blueshift of up to around 20 meV across the entire power range was noted from the free IX emission as the incident power density increased. This can be explained by the enhanced repulsive interactions between IXs with a dense exciton concentration. The repulsive dipole-dipole interactions are indicated by the blueshift in the experiment[63,100] and have been reported to intensively influence the diffusion of the excitons[57,124-127], due to its long-lived feature.

The high density of IXs enables their efficient transport across the heterostructures (Fig. 2b)[57,85,124] with diffusion length of micrometre-scale. This stems from the strong repulsive exciton–exciton interaction between the oriented dipoles, which becomes even manifest at higher exciton densities. The Stark shifts of hybrid excitons have also been reported with applied electric fields because of their interlayer character[17,84,128]. However, the control capability is closely related to the degree of hybridization, which can be determined by the interplay between Coulombic dipolar repulsion and attractive exchange interaction and has been proven to give rise to variable effective dipole lengths in the hybridized states, enabling the modification of many-body interactions and effective tuning of exciton transport[53,129].

### 3.1.1 Moiré exciton

The capability of creation of a moiré superlattice in stacked 2D semiconductors by varying mutual orientation of the monolayers introduces the emission of moiré excitons. They are different from the free IXs with a relatively uniform spatial distribution and stem from the moiré trapped states with spatially varying potentials[48,90], distinct from disorder-localized excitons[130,131] that arise due to imperfections or disorder in the crystal lattice with a confined spatial distribution but limited to regions where disorder or defects are present. Three distinct local minima with high-symmetry stacking configurations were formed within a moiré supercell, where excitons could be trapped. They arise from the registry of local atomic arrangements within the constituent crystal structures and were denoted[46,48,95,132] as $R_h^h$, $R_h^x$ and $R_h^M$ in the zero twist supercell, referring to different alignments as indicated in Fig. 2c. The moiré pattern induces an intriguing effect of lateral position-dependent modulation in the optical selection rules across the superlattice[46,133,134]. This is considered as a signature of moiré excitons (Note S4). Depending on the stacking alignment (near 0° *R*-type or near 60° *H*-type) between two monolayers, the energy landscape for interlayer excitons can be modified by variable depths of the moiré potential.

The calculated moiré potentials for two types of superlattices are illustrated in Fig. 2d, suggesting that the potential in 0° is much deeper than that in 60°, which is supported by the stronger spatial variations at 0°[48]. In addition, *R*-type stacking has an additional subtrap next to the deep potential, while *H*-type creates only a shallow potential. This moiré landscape creates twist-angle-dependent energy barriers for IXs, and is therefore a useful tool for manipulating the transport of them. This transforms the long-range diffusion of IX, similar to that in a coupled quantum well, into a tunable and controllable unique property. Compared with the normal diffusion behaviour of the monolayers (around hundreds nm[135]), anomalous diffusion was observed for IXs from both 0° and 60° stacking (Fig. 2e), presenting a much longer travelling distance and a substantially faster migration speed, which can be explained by the dipolar repulsion[18] as explained above. However, faster transport in the 60° heterostructure was demonstrated because the excitons were more mobile at 60° than at 0° under the same conditions[48]. The enhanced speed of exciton migration in the 60° heterobilayer provides additional compelling evidence for the fewer barriers the excitons would encounter when they move from one location to another with shallower potentials.

Moreover, the spatially and spectrally resolved PL from three $MoSe_2$/$WSe_2$ samples with different angles indicate that the diffusion of IXs is highly controllable, depending on the geometry of the moiré period (Note S5)[136,137]. The diffusion lengths of IX can be decreased from ~16 µm to ~1 µm or even 0 simply due to trapping at the moiré sites[136]. The dipole-dipole interaction plays a crucial role in accelerating the transport of IXs but the moiré pattern impedes IXs diffusion in the heterobilayer. Moiré IXs may serve as tunable sources for the transport of IXs and circularly polarised photons. It's worth noting that the interplay between dipolar repulsion and moiré trapping significantly governs exciton dynamics, which can lead to the delocalization of moiré excitons across the moiré potential[138].

### 3.2 Dipolar interlayer charged exciton

Indirect trions, formed from interlayer coupling and similar to IXs, also known as charged IXs, are dipolar particles consisting of two electrons (holes) and a hole (electron) confined in the separated layers. The characteristics of the interlayer trion state from van der Waals heterostructures were proposed and theoretically predicted by *Thygesen et al.*[86].

Subsequently, the experimental results confirmed the presence of bound indirect trions beneath the neutral IXs[62] (Fig. 3a). They were observed in non-intentionally doped MoSe$_2$/WSe$_2$ heterostructures and originated from the binding of electrons and holes generated through excitation to background charge carriers existing in the heterostructures. The measured binding energy for interlayer trions is approximately 26-28 meV, consistent with the calculated value[86], but another study reported a binding energy twice as small[124]. These values are smaller than the intralayer trion binding energies of around 30-40 meV[88] (Table 2) due to the spatially separated electrons and holes[22,86,139]. They are also smaller than that of IXs, stabilizing the neutral IX system and preventing their transformations into trions and charged particles[62].

Interlayer trions exhibit analogous behaviour to their direct counterparts as the temperature is varied[140,141] (Note S6). However, the electric tunability of the peak energy of interlayer trions was demonstrated due to the presence of a dipole moment, which allows for more flexibility even compared to interlayer neutral excitons because they not only have a permanent dipole but also possess a net charge and nonzero spin[59]. The indirect luminescence, interlayer excitons and trions, both exhibit an energy shift with the voltage-induced electric field, but the intralayer luminescence including trions is not controlled by the applied voltage (Fig. S3b) due to a lack of a built-in dipole oriented in the direction of the applied voltage[62,142,143].

The interlayer trion can be also formed by modifying the electrostatic conditions in heterostructures by introducing free charge carriers to the TMD layers, while maintaining a constant electric field across the TMD heterostructure[124]. This can be achieved by using a dual-gate device that allows the injection of carriers without applying an electric field to the heterobilayer[144]. The sudden redshifts for the n-doping and p-doping of MoSe$_2$/WSe$_2$ in Fig. 3b indicate the formation of negative and positive trions from neutral excitons, respectively. The redshift continues with the increasing electron or hole densities due to the enhanced doping level, which follows the intrinsic trion dynamics[140,145] and corresponding behaviours with the changing of gate voltages. This should be distinguished from the redshift originating from the Stark effect because the electrical field is fixed here. It is worth noting that trions can be referred[124] to as attractive polarons[146,147] (excitonic states interact attractively with a polarized fermionic sea), experiencing attractive interactions, consistent with the calculations illustrated above in Fig. 1f[22].

### 3.2.1 Charged moiré exciton

Similar to the moiré exciton, the interlayer trion can be confined to the potential to form a moiré-trapped trion. The reduced excitation power can effectively decrease the exciton density, which weakens the repulsive exciton–exciton interaction between exciton states and thus improves the modulation of the moiré potential on them, driving the behaviour of excitons from free particles to localised features. Their emissions then display fine, discrete, and closely spaced lines in the PL map (Fig. 3c). The peak splitting is caused by trapping through the moiré potential, and energy separation is closely related to the energy difference between the inter-minibands[144]. The binding energy of moiré trions (~7 meV) observed from a *H*-stacked heterobilayer has been found to be smaller than that of free interlayer trions, because the confinement imposed by the moiré potential restricted the wavefunction's variation, preventing it from achieving maximum binding[56].

Landé g-factors have been shown[47,55] as a unique identifier for discerning excitons confined within moiré potentials from those bound to atomic defects. But due to the smooth trapping potential in *H*-stacked heterostructures, the moiré states including excitons and trions inherit the *g*-factors of free IXs[56]. All observed moiré states exhibit a valley-Zeeman splitting with sharp emissions because the peak energies with $\sigma^-$ and $\sigma^+$ polarized light experiencing an opposite shift in the presence of a magnetic field. While the inhomogeneity of moiré traps results in a distribution of peak energies for IXs and charged IXs, the almost same energy shift with magnetic field is observed across all the moiré emitters, which is indicative of moiré trapped states[47]. This effect of trion energy can be viewed as the combined influence of the Zeeman shift of a valley exciton and that of the additional carrier. The extra electron/hole in a trion solely alters the transition energy without contributing to the spectroscopic Zeeman splitting[56]. Intriguingly, a transition from co-circular to cross-circular polarizations of moiré trions is demonstrated when it shifts from negatively to positively charged, which arises from the interplay between valley-flip and spin-flip energy relaxation processes of photo-excited electrons during the creation of interlayer trions[56].

Interlayer moiré states showed an interesting dependence on carrier densities. Unlike the modulation observed in free IXs with varying carrier densities, the energy of moiré excitons remains nearly constant as charge densities increase. This distinct behaviour stems from the distinct nature of exciton–carrier interactions: IX states can freely move through the sample, interacting with ambient carriers to produce an appreciable energy shift, while moiré exciton states are localised in the trap, only weakly interacting with some carriers outside the moiré cell[144], unless increased exciton-exciton repulsion leads to a shallower potential due to enhanced exciton density[138]. Moiré trion emission exhibits a more striking carrier density dependence, appearing on one side but vanishing on the other with varying doping levels (Fig. 3c). Stark shifts in moiré trion energies are observable with out-of-plane electric fields (Fig. 3d), despite being trapped in the potential[56]. Intensity plots (Fig. 3e) of moiré trions from an *H*-stacked MoSe$_2$/WSe$_2$ sample at different temperatures reveal visible PL at lower temperatures (approximately 2 K), diminishing rapidly above 10 K due to the thermal delocalization of moiré trions in the shallow potentials of these heterobilayers.

Furthermore, moiré trions can also interact with charge carriers trapped in neighbouring moiré sites through Coulomb interaction, modifying the energy emissions of moiré trions[148].

The Coulomb interaction energy between a trapped interlayer trion and a single nearby trapped electron was modelled and calculated as a function of moiré period (s) (Fig. 3f) based on the small interlayer distance ($d \ll s$). It decreases at a rate of approximately $1/s^2$ and may cause staircase-like energy jumps in moiré trion energy[148] instead of a continuous and smooth shift.

### 3.3 Dipolar interlayer biexciton

A robust dipolar biexciton is a strongly bound state of two interlayer excitons, which was predicted to form only when the interlayer distance ($d$) is sufficiently small in 2D systems (Fig. 4a)[41,112]. It aligns with the theoretical studies of exciton-exciton interactions[34] explained above (Fig. 1e and f), which identified the critical layer separation[22,105]. This certain threshold makes the observation of free interlayer biexcitons more challenging compared to intralayer biexcitons, owning to the classical dipole-dipole repulsion among IXs, in addition to reduced Coulombic attraction and weaker binding energies[64,149-151] The separation distance between the two layers affects the coupling[151] and effective interlayer charge transfer[152-154], and considerably influences the interactions between the dipoles.

The suspended atomically thin TMD hetero-bilayer of $WSe_2/WS_2$ fulfills the required condition ($d < 0.87a_B$) and has strongly enhanced dipole-dipole interactions, providing a perfect platform for the first experimental observation of the appearance of dipolar biexciton phases[64]. Interlayer biexcitons were observed in an $H$-stacked hetero-bilayer with moiré-trapped excitons. The distinct energy shifts (Fig. 4b) with varying exciton densities demonstrated that they were governed by opposite interactions, revealing the transition from repulsive to attractive, and suggested the contribution of interlayer biexcitons. This was confirmed by a signature property of biexciton emissions, showing a super-linear relationship between integrated PL intensities and excitation power (Fig. 4c). At the same time, the attractive interactions among IXs give rise to the negative binding energy of this interlayer biexciton complex. This is different from the biexcitons observed from a potential trap[54,63] such as localized interlayer biexciton or moiré biexciton with a positive binding energy.

### 3.3.1 Localized biexciton

In contrast, another type of a biexciton complex can be formed more easily and passively by localizing exciton states in traps. They are generated because two IXs are confined to specific regions or moiré traps within the heterostructure. This results in finite positive binding energies, and the interactions between them remain mainly repulsive, while free interlayer biexcitons are formed by spontaneously and attractively binding two IXs together. The localized biexciton was experimentally distinguished in an hBN-encapsulated $MoSe_2$/$WSe_2$ hetero-bilayer (Fig. 4d). Instead of a hydrogen molecule-like bound state, this observed biexciton was a double-occupancy state of a localised trap[54]. It has a significantly smaller dipole moment and requires tighter localization, and the two excitons are correlated owing to repulsive interactions. In comparison to their free and intralayer counterparts, the biexciton states emit at lower energy side with respect to the exciton[64,150,155,156], whereas the emission of localized interlayer biexcitons is blue-shifted from the interlayer exciton (Fig. 4d). The emission energy of this interlayer biexciton was raised to a higher energy state by the dipole-dipole interaction, as it required on-site energy to overcome the dipolar repulsion for generating two IXs within the same trap (Fig. 4e).

The potential with a lateral size larger than the Bohr radius of interlayer exciton allows interlayer excitons to be trapped in, which can be formed due to the moiré superlattice[157] or possibly caused by strain or defects. Similar trapped biexcitons and higher-order states were observed[63] (Fig. 4f) from the nanoscale confinement potential induced by placing the sample onto an array of nanopillars (Fig. 4f, inset). The correlated multiexcitons were studied to suggest that the significant dipolar and exchange interactions between IXs can provide a high possibility for the realisation of multi-particle high-correlation phenomena in layered van der Waals heterostructures[63].

**3.4 Exciton condensation and correlation**

The strong correlation between dipolar excitons is mainly reflected by dipole-dipole interactions, which substantially influence the behaviour of the entire exciton system and thus affect the potential properties of semiconductor devices. The concept of Bose-Einstein condensation is understood as a coherent state of a gas of bosons when the temperature drops very close to absolute zero[158,159]. At this extremely low temperature, the wave size of each particle becomes larger than the average interatomic distance and all bosons can be described as having the same energy in the same quantum state. The quantum coherence of bosons is one of the most brilliant macroscopic manifestations of the quantum mechanical law in the microscopic world. After the discovery of Bose-Einstein condensation in atomic vapours[160], excitons have received much attention due to their potential for observing the condensation of bosonic quasiparticles. Exciton systems are characterised by strong exciton correlations based on the environmental temperature and exciton concentration[28,41].

3.4.1 Phase transitions of dipolar exciton system

A general picture of exciton-exciton correlations was qualitatively estimated with these two crucial parameters in coupled quantum wells[28], demonstrating the different phases according to the correlations occurring in the system (Fig. 5a). Notably, the system behaves as a gas (region II) when the exciton density is low and has a strong pair correlation. However, correlations are not applicable if the temperature is high and the mean field approximation dominates the system instead. Further reduction in temperature at lower exciton densities brings the system from region II to III, resulting in the quantisation of the exciton-exciton scattering and inducing degeneracy of the system, accompanied by strong multiparticle correlations. Thus, the system is considered a Bose gas rather than a classical gas in this region, and the mean free path no longer exists in the system. At high concentrations, quantum degeneracy was suppressed by the strong confinement of the wave function of each exciton because of the dipolar repulsion between them. Therefore, the system exhibits a classical liquid in some regions, where the probability of finding more than two excitons close to each other and becoming an order of the unity, the contact interaction does not exist, and the correlation is from multiexcitons.

A similar phase diagram was later developed using TMD stacked monolayers to elevate the temperature for achieving the degenerate state of IXs to record-high levels (Fig. S4a)[41]. More interestingly, when the wave functions of the excitons overlap, it competes with the dipole-dipole repulsive force; an enhancement in concentration directly induces the Mott transition, and there is no classical liquid behaviour. The larger effective mass of IXs in TMD-based heterostructures (Table 1) may seem unfavourable for attaining a higher critical temperature, based on the usual inverse relationship between them[161-163]. However, the opposite holds true when exciton density is under control[41], and the twisted TMD bilayers provide a fascinating method to finely tune and enhance this. Thanks to their substantial exciton binding energy, IXs in TMD bilayers exhibit stability against thermal dissociation well beyond room temperature across a wide range of electron densities.

### 3.4.2 High temperature exciton condensation

In the correlated liquid regime, the excitons can act as superfluids. This high-temperature superfluidity has been predicted in TMD heterostructures[41] and was also anticipated in other systems like correlated-electron systems[161], revealing that the origin of high-temperature superconductivity lies in strong electron correlation. The high degree of angle tunability provided by TMD heterobilayers enables precise control of this correlation through exciton density. An exotic quantum state, known as pair superfluids[107,164] was introduced to distinguish the superfluidity of interlayer bound pairs from the independent dipole condensation of each layer. The interlayer separation distance ($d$) again is very important here, as it significantly influences the correlated phases in the exciton system because of unique representative interactions. The independent superfluids at distinct layers are normally performed for the large interlayer separation distance[165,166], because the layers are decoupled or minimally coupled, while when the distance between two layers decreased to a limit $d \ll a_B$, the intense attractive component in dipolar interactions could generate the dipolar bound pairs[167,168].

Atomically thin 2D heterostructures have been recently recognized as optimal platforms[169] for high-temperature excitonic condensation, surpassing the potential of previously explored monolayer 1T-TiSe$_2$, where Peierls instability impedes excitonic condensation[169-172]. A threshold behaviour of the electroluminescence from 2D van der Waals semiconductors was experimentally observed to support the exciton quasi-condensation[23]. The integrated intensity of electroluminescence emission exhibited a critical threshold dependence on the exciton density (Fig. 5b), which increased anomalously by around two orders of magnitude around the threshold density, while the tunnelling current only raised by a factor of 2. The phenomena survived above around 100 K, which was consistent with the predicted condensation temperature of the IXs in the van der Waals hetero-stack system[42,173].

Behaviour in good agreement with condensation was also reported for photogenerated excitons in the MoSe$_2$/WSe$_2$ heterostructure, indicating the degeneracy of the many-body states of IXs[24]. Below the expected critical degeneracy temperature, the coherence of interlayer excitons becomes clearly observable with a significantly reduced but consistent spectral linewidth in PL emissions. Similar behaviour also emerges when the excitation density exceeds the specific critical point, and it validates the presence of strongly correlated many-body states in the system (Fig. 5c-f and Note S7). The observations provide evidence and opportunity for exciton condensation at high temperatures because the threshold behaviour with respect to temperature and density is indicative of the degenerate many-body exciton state. These experimental realisations of exciton quasi-condensation and quantum degeneracy fall within region IV of the phase diagram (Fig. 5a).

4. Outlook

The remarkable characteristics of interlayer excitons in TMD bilayers make them a promising candidate for optoelectronic devices (Note S8). Considering excitons in TMD heterostructures as either a localised, quantum-confined system, or a many-body interacting system holds benefits for future device applications[14], but a regime that can effectively combine these two limits is of paramount importance and still needs to be investigated. The study of IXs in stacked 2D heterobilayers offers an interdisciplinary opportunity to enhance future device performance.

We envision five different areas, where dipolar many-body complexes will be the focus of research, including quantum phase transitions in moiré materials involving IXs, interfacial charge transfer (CT) excitons in lateral TMD heterostructures, Wigner crystallization of IXs, quantum light emission of interlayer biexcitons as well as strategies for enhancing the intrinsically low emission intensity of IXs. In the following, we briefly discuss these different areas and their potential.

From the perspective of physics, the IXs in moiré patterns provide a paradigm to investigate Bose-Hubbard model physics, including the quantum phase transition between the Mott and extended exciton phases[174,175]. Although initial studies have been performed, this fascinating research area is still in its infancy stage. The emphasis on the observation of many-body correlated complexes serves a dual purpose, discovering more correlated quantum states and achieving potential breakthrough developments in future real life (Note S9).

The emerging research field of lateral TMD heterostructures and the appearance of CT excitons has recently attracted attention [176-178]. The spatial separation of CT excitons results in an in-plane dipole that is typically much larger than for interlayer excitons in vertical heterostructures, where the dipole is limited by layer separation. The progress in growth techniques allows for the production of high-quality ultrathin interfaces, facilitating the formation of stable and highly dipolar CT excitons that have a high potential to boost exciton transport and exciton dissociation, both highly relevant for optoelectronic applications. So far, little is known about the formation dynamics of CTs excitons nor their transport across and along the interface, and this needs to be investigated in the future.

### 4.1 Wigner crystallization of IXs

In the regimes of strong coupling for IXs, crystal phases become energetically favourable, allowing access to interlayer exciton solids even at lower exciton concentrations[107]. Dipolar crystals consisting of indirect excitons have been reported in semiconductor quantum wells[179] and heterobilayers[59,180]. Strongly correlated collective interlayer trions are expected in the highly excited van der Waals heterostructure, including the crystallisation of the unlike-charge interlayer trions and Wigner crystallisation of the like-charge interlayer trions. This can be selectively realised with a properly chosen effective mass of electrons and holes in TMD bilayers by simply varying the separation distance (*d*) between layers in addition to using electrostatic doping[59]. Another scenario that can induce Wigner crystallization of interlayer trions is a strong external magnetostatic field[181,182], perpendicular to the TMD bilayer structure. Under this condition, interlayer trions can be trapped by the parabolic potential induced by the magnetic field, which stimulates crystallisation owing to the Coulomb interaction between them[180]. It has been reported that the superfluids in the IX system can be transformed into crystals upon compression for indirect excitons with a large e-h separation due to the strong exciton-exciton interaction, and the second transition between solid and liquid is expected with the changing in temperature and density[179].

The initial studies have left many questions open and exploring more correlated and complex dipolar interactions can open up new avenues for diverse quantum discoveries. Composite particles involving three, four, or more constituent particles could offer insights into pursuing room-temperature superconductivity[183,184] as these highly correlated systems manifest much stronger interactions. Therefore, there has been ongoing research on higher-order correlated entities, including theorised four-body entities[185] and other complex composites consisting of even more body states but also here further experimental and theoretical studies are needed, as this will facilitate a better understanding of the structure of matter and interactions or correlations among systems of constituent elementary particles.

### 4.2 Quantum light emission of dipolar biexcitons

Building on the early successes of four-body clusters of biexcitons, there is great potential for discoveries and applications in quantum optics. Generating entanglement between quantum photons is a central goal in quantum information science and is significantly beneficial for quantum communication and technology[186-189]. Atomically thin semiconductors are suitable material platforms that offer quantum interfaces in scalable and compact devices. The quantum states of photons from a material can be tightly connected to quantum emitters through light-matter interactions[157]. The photoluminescence emission process of biexciton in semiconductor nanocrystals has been experimentally proven to have a cascaded nature, which leads to an emission cascade of single photons[155] and is actually a sequential two-photon emission process[190]. Essentially, the biexciton emission can be understood as a pair of entangled photons, meaning that these two photons are closely interrelated and may have a single wave function. However, fine structure splitting[155] due to electron-hole exchange interactions limits the entanglement fidelity for classical intralayer biexcitons. The capability of biexcitons to emit quantum light was confirmed[155] by the second-order photon correlation measurement $g^{(2)}$, which verified the harnessing of biexciton emissions as a single-photon quantum light source.

Highly entangled photons with fidelities larger than 0.99 are preferred to serve as fundamental building blocks for implementing quantum communication and computation[191]. The interlayer biexcitons with electrons and holes entitled on different layers have the potential to achieve this ideal entanglement fidelity and will certainly be an important topic of future research. Principally, the exchange effect of electrons and holes is absent in this vertically separated structure, which is devoid of fine structure splitting and is not subject to limitation. The tunability of interlayer states enables them to function as tunable quantum light emitters with the capability of adjusting the energy range through an external electric field. This has stimulated further exploration and development of other dipolar complexes.

### 4.3 Improvement of interlayer emission

Finally, the emission intensity of IXs depends heavily on interlayer coupling. Restricted transitions limit exciton-photon interactions, and their long lifetime and transport distance moderately influence the emission rate, weakening the optical response of IXs. Some discussed critical factors, such as twist angle and dielectric screening, may appear challenging to control in real devices. However, recent proposals suggest potential methods to overcome these limitations. A groundbreaking robotic four-dimensional pixel assembly technique has been introduced[192] with unprecedented speed, intentional design, and extensive control over both area and angles, to address challenges in assembling structures from micromechanically exfoliated flakes for scalable and rapid manufacturing. This automated assembly streamlines structure construction, rendering it applicable to practical device manufacturing. Additionally, protecting TMD materials during large-scale fabrication involves the use of low dielectric materials, such as certain low dielectric polymers or composite materials[193], which exhibit improved scalability and optical performance with lower dielectric screening. Advancements in low dielectric materials provide a promising outlook for achieving an ultra-low dielectric constant in manufacturing[194]. This marks a significant progress in shifting from TMD-based devices with limited functional areas to the development of large-area devices in a dielectric environment. These hold the potential to fully unlock the capabilities of TMD heterostructures as a versatile platform for practical and scalable applications.

On the other hand, single-photon emissions from interlayer biexcitons are challenging to observe due to their weak intensity, hindering entangled photon detection[54]. This is expected to be resolved in future studies by incorporating plasmonic nanocavities or metasurface microcavities, which have demonstrated effective control over radiative properties of monolayer quantum emitters[195-197] and modulation of the out-of-plane emission in layered semiconductor flakes[198] and dark excitons[199-201]. Emitter–plasmon structures can increase the intensity and emission rate, achieving pronounced optical nonlinearities and improving emitter quantum yields. Similar to dark excitons, the out-of-plane dipole moment in the optical transition of IXs makes them efficiently enhanced through coupling with plasmonic nanocavities. The TMD-plasmonic device can excite surface plasmon polaritons (SPP) that are strongly polarised in the out-of-plane direction[199]. Thus, the induced emission rate for an out-of-plane dipole is much greater than that for an in-plane dipole[201] when the radiative emission is coupled with the SPP. This should facilitate the remarkable enhancement of interlayer exciton emissions owing to the giant coupling between the IX dipole transition and the induced electric field.

Currently, cavities are recognized as essential tools to manipulate the optical performance of ultra-thin semiconductors because they can concentrate light and confine photons to enhance light-matter interactions, enabling serval applications that require microscopic volume confinement. Another quasiparticle known as interlayer exciton-polariton, can be produced as superpositions of excitons and photons under the strong light-matter coupling in a cavity, with significantly enhanced non-linearity[202-204]. Other effective means of improving the emission efficiency of IXs require further exploration, and different nano-fabrication techniques can be considered for incorporation during the following step, which will activate more promising discoveries and create novel nano-optoelectronics devices in future studies.

Overall, stacked 2D semiconductor structures have displayed significant potential for achieving a diverse range of advanced physical phenomena, presenting a powerful and versatile system that allows device integration with sophisticated quantum optical performance. TMD hetero-bilayers offer the possibilities of compact quantum circuits and scalable device fabrication. We envision that the demonstrated interlayer complexes, such as biexcitons, serve as ideal building blocks for functional quantum devices, and the static dipole moment in dipolar structure enables natural ways to steadily realise novel many-body states. Additional discoveries regarding 2D hetero-bilayers will be inspired and expanded by further investigations of other dipolar complexes, such as CT excitons in lateral TMD heterostructures. The dipolar IXs arising from anisotropic 2D materials[205-210], such as black-phosphorus, organics, and 2D perovskites, could be intriguing to explore. Their unique properties lie in facilitating innovative device applications across various wavelength ranges, including the mid-infrared and even the far-infrared spectrum. The continuing progress of nanofabrication techniques and advancement of vdWs heterostructures will drive the future growth of the field, providing access to a broader spectrum of physics and more applications over a wide range. The complete integration of individual excitonic devices based on dipolar systems that cover diverse quantum effects on a single chip is highly possible in the near future.

**Acknowledgements**

The authors would like to acknowledge funding support from the ANU PhD student scholarship, Australian Research Council (grant No.: DP220102219, DP180103238, LE200100032), ARC Centre of Excellence in Quantum Computation and Communication Technology (project number CE170100012) and National Heart Foundation (ARIES ID: 35852).

**Author contributions**

Y. L. and X. S. conceived the study. All authors contributed to the writing and all aspects of the manuscript and were supervised by Y. L.

**Competing interests**

The authors declare no competing interests.

**Key points**

- Stacked 2D heterobilayers host many-body excitonic states with strong dipole-dipole interactions, enabling the investigation of multi-particle correlations and novel quantum physics.
- The physics of dipole-dipole interactions in 2D heterobilayers is at the origin of several important phenomena, including Bose-Einstein condensation (BEC), superfluidity and supersolids.
- Long-range dipolar interactions and the distinctive tunability of 2D heterobilayers offer a promising platform for highly controllable studies through various means, such as interlayer distance, excitation densities, external fields, environmental stimuli, and important moiré superlattice.

- The strong binding energy of dipolar complexes in 2D stacked heterobilayers allows for the pursuit of room-temperature condensation and superfluidity.
- The impact of dipolar many-body complexes extends beyond excitonic physics discoveries and forms the basis of several key concepts, including Wigner crystallization, tunable quantum light emitters, and scalable quantum devices.

| Exciton properties | IX | IX |
|---|---|---|
| Material platform | Coupled quantum wells | TMD heterobilayers |
| Effective masses | 0.22-0.40 $m_0$ | 0.9-1.1 $m_0$ |
| Interlayer distance | ~12 nm | ~0.7 nm |
| Bohr radius | ~10 nm | ~0.93 nm |
| Binding energies | ~4-10 meV | ~100-300 meV |
| Diffusion length | Scale of μm | Scale of μm |
| Lifetime | Up to tens of μs | Up to few μs |
| Critical transition temperature | < 10 K | >100 K |
| Ref. | 34,35,110,125,211-216 | 20,23,41,64,65,124,217,218 |

**Table 1.** │ Exciton properties of IXs from conventional double quantum well structures and 2D TMD heterobilayers

| Exciton species | Binding energy | Ref. |
|---|---|---|
| **X** | ~500-1000 meV | 65,149,219,220 |
| **X$^{-/+}$** | ~30-40 meV | 88,149,221,222 |
| **XX** | ~50-60 meV | 149,221,223,224 |
| **IX** | ~100-300 meV | 65,217,225 |
| **IX$^{-/+}$** | ~10-28 meV | 22,62,86,124 |
| **IXX** | ~15-18 meV | 64 |
| **Moiré IX** | ~400 meV | 51,226 |
| **Moiré IX$^{-/+}$** | ~7 meV | 56 |
| **Localized IXX** | ~1-15 meV | 54,63 |

**Table 2.** │ Binding energies of exciton complexes in TMD monolayers and heterostructures.

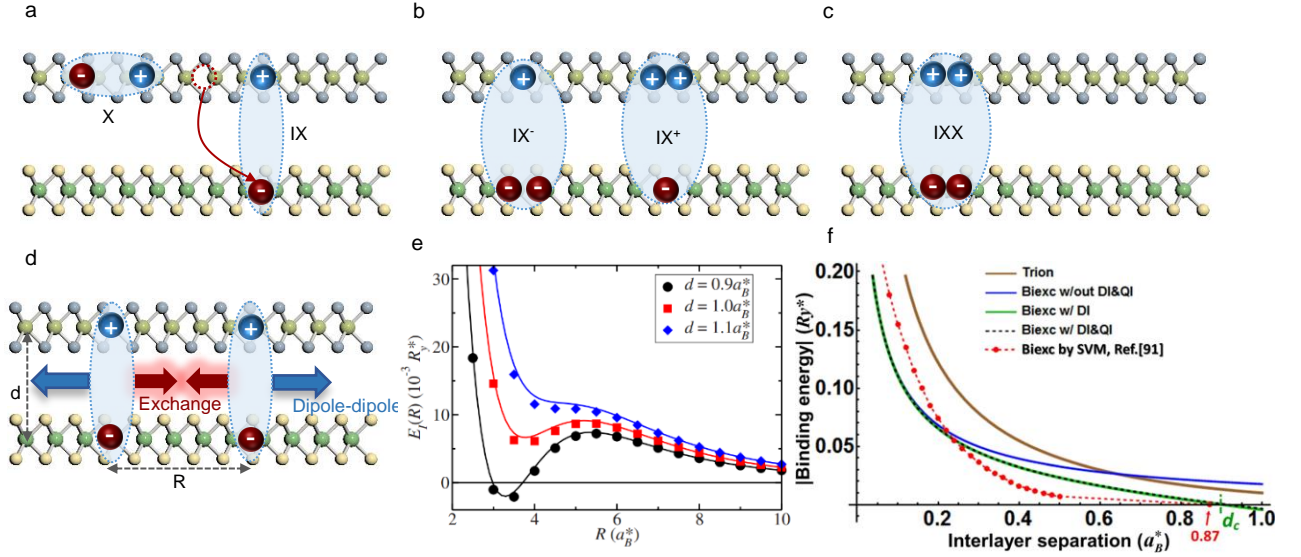

**Fig 1. │ Interlayer exciton and multi-complexes in semiconducting bi-layers. a,** Schematic illustration of intralayer exciton and interlayer exciton in the real-space side view of stacked heterobilayers. Red arrow indicates the electrons or holes tunnelling from one monolayer into the opposite, forming the interlayer structure. **b,** Illustration of interlayer negative trion and positive trion. **c,** Illustration of interlayer biexciton, consisting of two interlayer excitons. **d,** Schematic illustration of dipole-dipole interactions between dipolar excitons in the double layers separated by a distance d. Blue and red arrows represent the dipole-dipole repulsive and attractive interactions respectively, and R indicates the distance between excitons. **e,** Calculated dipole-dipole interactions as a function of exciton separation R for different interlayer separation distances. The black solid line shows a transition point of interaction from repulsive to attractive. **f,** Calculated binding energies of interlayer biexciton and trion complexes as a function of the interlayer separation distance, considering with and without the long-range interactions, pointing out a critical threshold value of interlayer distance for the formation of biexciton. Adapted with permission from: **a**, ref. [101], Springer Nature Ltd. Reproduced with permission from: **e**, ref. [105], American Physical Society; **f**, ref. [22], American Physical Society.

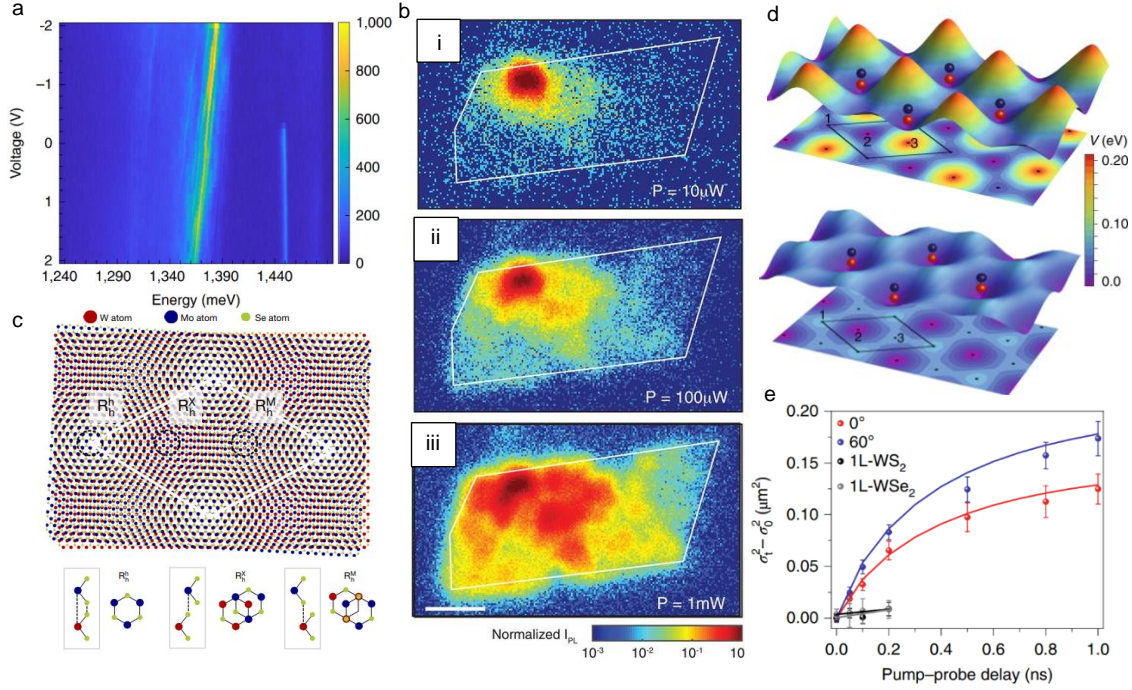

**Fig 2. | Dipolar repulsive interaction and modulation in carrier diffusion. a**, PL mapping of interlayer and intralayer excitons from MoSe$_2$/WSe$_2$ heterobilayer with the electric field tuning. The interlayer exciton exhibits a redshift of around 20 meV, confirming the existence of an out-of-plane dipole, but intralayer exciton does not show E tuneability with the changed of gate voltage. **b**, Intensity-normalized spatial PL mapping from MoSe$_2$/WSe$_2$ heterostructure under different excitation powers (10 µW to 1mW) at 4 K. The white solid lines indicate the shape of heterostructure. The continuous wave laser excitation is fixed at the top left of the sample. **c**, Moiré superlattice formed from a *R*-type MoSe$_2$/WSe$_2$ hetero-bilayer. Three different local atomic registries ($R_h^h, R_h^x, R_h^M$) generated are highlighted in the black circles. **d**, The 3D illustrations and corresponding 2D projections of the K–K moiré potentials in *R*- (top) and *H*-type (bottom) stacking to show the traps formed of at local minima. 1, 2 and 3 indicate $R_h^h$, $R_h^x$ and $R_h^M$ in 0° (top) and $H_h^h$, $H_h^x$ and $H_h^M$ in 60° (bottom). **e**, Measured diffusion of interlayer excitons from heterostructures with different twist angles using the pump and probe beams, showing the interplay between the moiré potentials and strong many-body interactions. The exciton diffusion in 1L-WSe$_2$ and 1L-WS$_2$ are presented for comparison. Reproduced with permission from: **a**, ref.[54], Springer Nature Ltd; **b**, ref.[124], The American Association for the Advancement of Science; **c**, ref.[46], Springer Nature Ltd; **d, e**, ref.[48], Springer Nature Ltd.

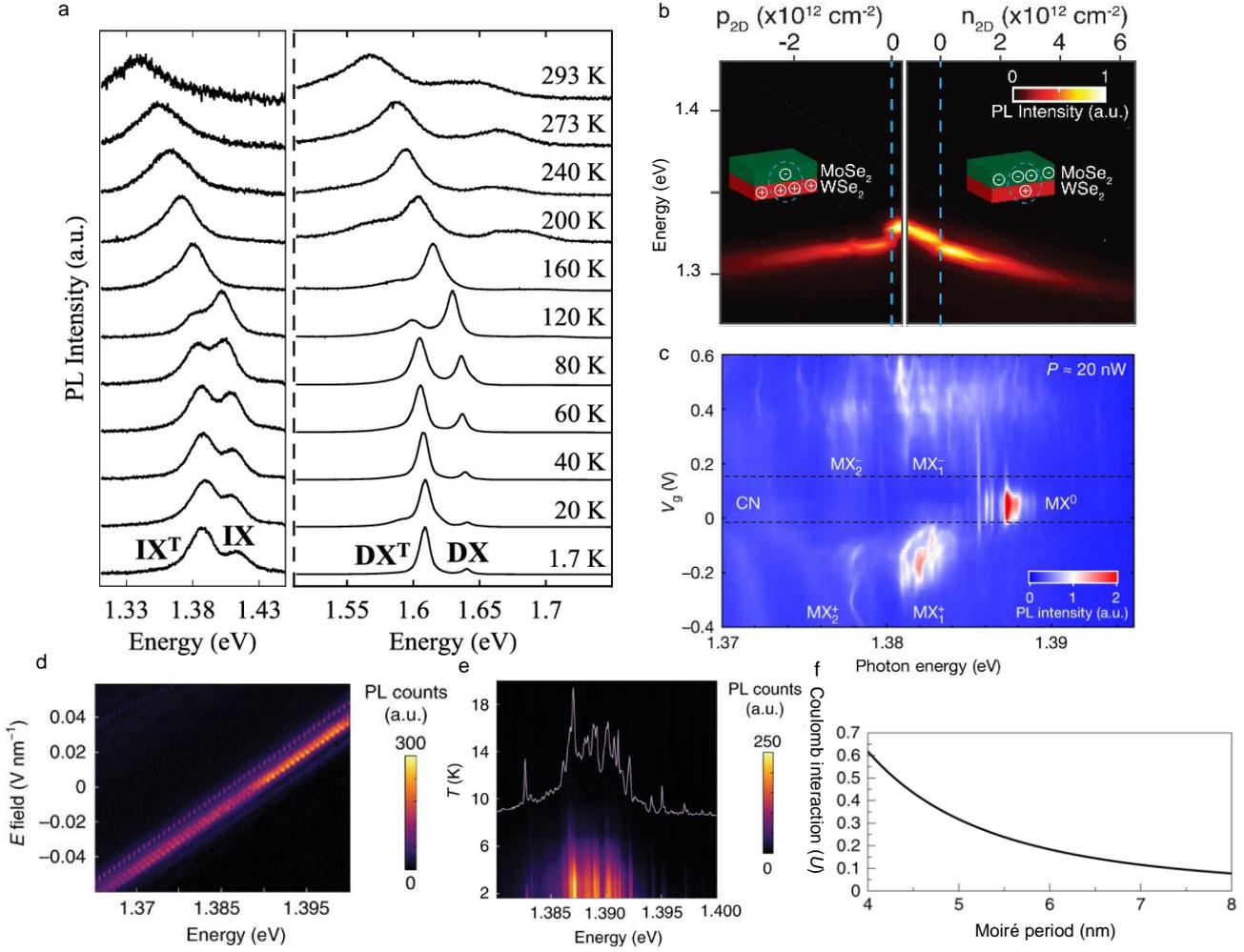

**Fig 3. │Dipolar interlayer trions. a,** Temperature-dependent PL spectra of interlayer (IX, $IX^T$) and intralayer (DX, $DX^T$) luminescence in $MoSe_2/WSe_2$ heterostructure. **b,** Carrier-density dependent PL shows the formation of charged interlayer excitons. The insets are the schematic illustration of hole-doped (left) and electron-doped (right) interlayer trion. **c,** Charge-density dependent PL maps of moiré excitons and trions. **d** PL spectrum of negative moiré trions as a function of applied out-of-plane electric field with fixed doping. **e,** Temperature-dependent PL intensity of positively charged moiré trions, showing the shallow potential of the moiré traps. Inset is the measured PL spectrum at 1.6 K. **f,** Calculated Coulomb interaction energy between moiré trion and a neighbouring single trapped electron as a function of moiré lattice spacing. Reproduced with permission from: **a**, ref.[62], American Chemical Society; **b**, ref.[124], The American Association for the Advancement of Science; **c**, ref.[144], Springer Nature Ltd; **d**, **e**, ref.[56], Springer Nature Ltd; **f**, ref.[148], Springer Nature Ltd.

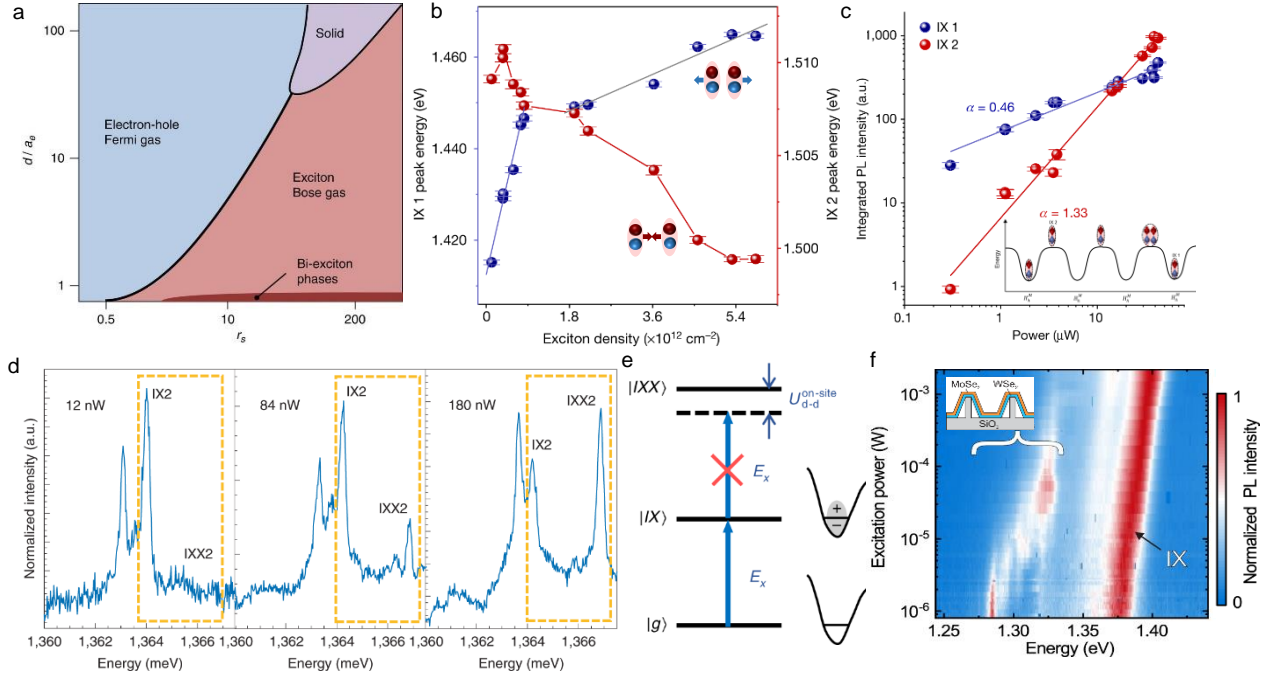

**Fig 4. │Interlayer biexciton and interactions. a,** Schematic phase diagram of an electron-hole system in the MoS$_2$/hBN heterostructure at T=0 K, indicating the formation of bi-exciton phases with a small interlayer separation distance. **b,** PL peak energy of interlayer exciton Peak 1 (blue balls) and Peak 2 (red balls) as a function of exciton density at 83 K. The blue shift is caused by the dipole-dipole repulsive interaction, while the red shift indicates an attractive interaction, leading to the appearance of biexciton phases. The blue and grey solid lines represent fitting curves using mean field approximation and exciton pair correlation models, respectively. **c,** Integrated PL intensity of interlayer exciton Peak 1 and Peak 2 as a function of pumping power. The fitting curves give rise to a sub-linear and super-linear relationship for peak 1 and 2, respectively, with the increase of excitation power. The inset is the illustration of the spatial variations of the moiré potential from the *H*-type stacked WSe$_2$/WS$_2$ heterobiayer, with the assignment of interlayer exciton states. **d,** PL spectra of interlayer exciton emissions from well-aligned (near 0°) MoSe$_2$/WSe$_2$ heterostructure under different excitation powers, ranging from 12 nW to 180 nW. The biexciton phase was highlighted in yellow dashed rectangles. **e,** Energy diagram of localized interlayer optical excitation in a potential well. |*g*⟩, |*IX*⟩ and |*IXX*⟩ denote the ground, single- and double-occupancy states, and it costs energy *Ex* for the transition from ground to single. The transition from exciton to biexciton cost an on-site energy due to the dipole-dipole repulsion. **f,** Normalized PL intensity spectra from MoSe$_2$/WSe$_2$ heterobilayer at 20 K with different excitation powers at a position on and outside the pillar. The discrete emissions including biexciton states are observed from the pillar region. Reproduced with permission from: **a**, ref. [41], Springer Nature Ltd; **d**, **e**, ref. [54], Springer Nature Ltd. Adapted with permission from: **b**, **c**, ref. [64], Springer Nature Ltd; **f**, ref. [63], Springer Nature Ltd.

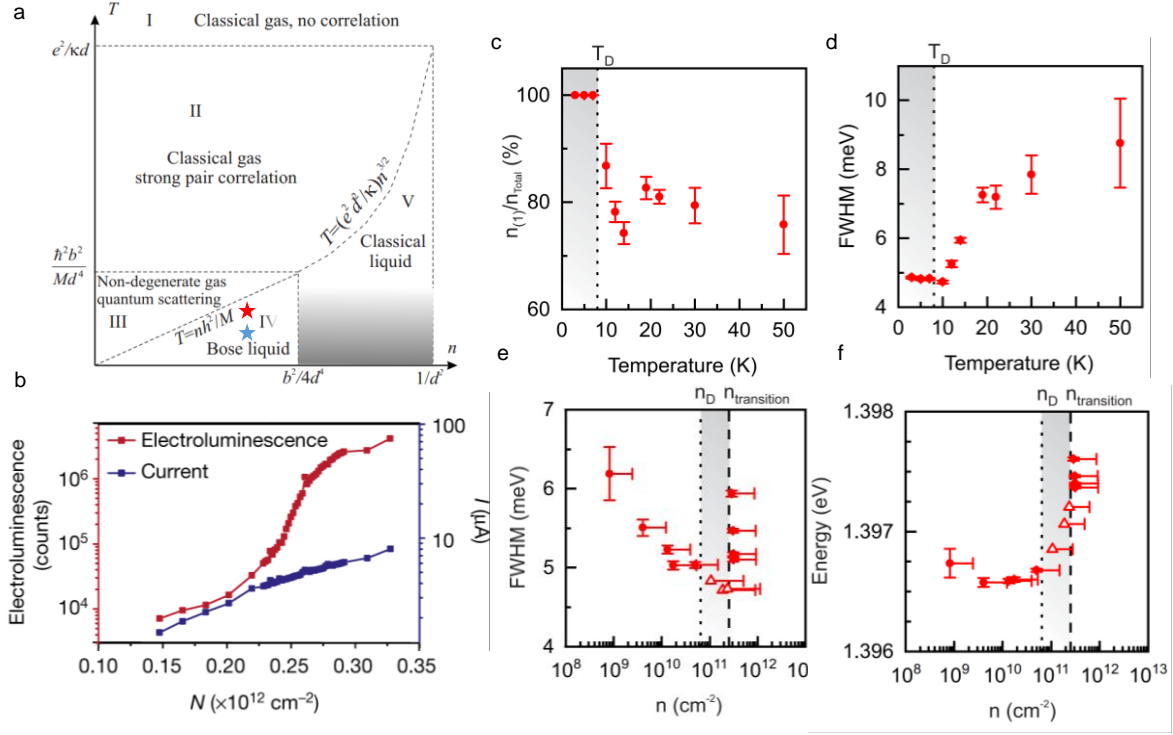

**Fig 5. │Exciton Condensation in atomic thin double layers. a,** Expected exciton-exciton correlations in the exciton system at the *n-T* plane. The red and blue stars indicate the corresponding positions of the experimental demonstration of exciton quasi-condensation and quantum degeneracy illustrated in **b** and **c-f**. **b,** Exciton electroluminescence and current as a function of exciton density in $MoSe_2$/2L-hBN/$WSe_2$ heterostructure**.** The electroluminescence spectra shows a threshold that is an indication of IX condensation**. c,** Estimated occupation of the manybody state as a function of temperatures, which shows almost 100% below the degeneracy temperature. **d,** Temperature dependent Full width at half maximum (FWHM) of many-body state. **e,** Photoluminescence FWHM of the many-body state as a function of exciton densities at T = 4 K. $n_D$ and $n_{transition}$ depict the degeneracy density and the transition of regimes, which implies the IX condensation. **f,** Emission peak energy of the many-body state as a function of exciton densities, $n_D$ and $n_{transition}$ are marked by the dotted lines. Adapted with permission from: **a**, ref. [28], American Physical Society. Reproduced with permission from: **b**, ref. [23], Springer Nature Ltd; **c**-**f**, ref. [24], American Physical Society.

**Box 1 | Formation efficiency of interlayer exciton**

The IX formation efficiency is closely related to the twist angle, temperature, and dielectric screening[64]. A schematic shown here illustrates how the efficiency could be maximised theoretically by optimising the three factors, as supported by previous reports and experimental data. Each of these three factors can be individually optimised to achieve a higher formation efficiency through different angles, whereas the efficiency can be maximised by combining them to work simultaneously in the system. Recently, the rotational angle between two monolayers in a hetero-bilayer has been reported as a critical parameter[227-230], attracting considerable interest and being important for modulating the optoelectronic properties of thin semiconductors. The interlayer exciton is caused by interlayer coupling and efficient charge transfer in the heterostructure; however, it is not observed in all heterobilayers, only in the finite area with effective coupling and good rotational alignment of TMD monolayers[227,228]. Van der Waals heterostructures allow for the formation without precise lattice matching and twist alignment. This characteristic enables the separation of electrons and holes in momentum space by introducing a momentum misalignment between the conduction and valence band edges of the two layers[91]. By adjusting the rotational angle, the recombination rate of IXs can be effectively tuned and for larger twist angles (> around 10°), the momentum mismatch becomes significant, which prohibits the radiative recombination although the interlayer coupling persists[228]. Thus, well-coupled heterobilayers are favourable for dipolar exciton studies.

Another important factor in maintaining the formation efficiency of IXs is a low operating temperature, as the photoluminescence (PL) intensity of IX is strongly temperature-dependent[231-233]. The typical TMD heterostructures can be identified as *R*-type or *H*-type according to the rotational angle being close to 0° or 60° due to the opposite band alignments of the layers in real and momentum space[48,233]. An increasing trend in the integrated PL intensity of IXs with decreasing temperature has been reported for both types of TMD heterostructures[233]. This observation validates the stronger IX emissions and higher IX formation efficiency at lower temperatures, regardless of the stacking arrangement of the heterobilayers.

In addition, low dielectric screening has also been experimentally proven to be the third key factor supporting sufficient IX formation efficiency. It has been found that hBN-encapsulation provides a strong dielectric screening effect and reduces the binding energy of IXs[22], which prevents the formation of IXs, and dielectric screening accordingly affects the interactions between IXs[102,234]. More different dielectric environments were designed to compare and experimentally characterise the interlayer exciton formation efficiency in TMD heterobilayers; the defined indicative parameter β shows a clear decline when the dielectric screening is enhanced (Fig. S1e). The theoretical prediction[217] was confirmed that reduced dielectric screening in heterostructures can result in a higher formation efficiency of IXs because of the strongly enhanced Coulomb interactions and broader excitonic wave functions in reciprocal space. Taken together, the three crucial parameters simultaneously play significant roles in the formation of interlayer excitons, which can be developed along different directions, but ultimately combined to reach the optimal conditions, efficiently generating desirable emissions and thus enhancing the dipolar interactions between IXs under a high level of exciton density.

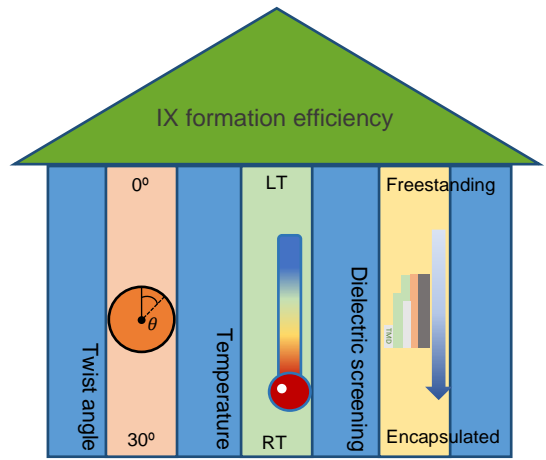

**Fig. Box 1:** Schematic illustration of achieving higher interlayer exciton formation efficiency by controlling three key parameters including small twist rotation, low operation temperature and weak dielectric screening effect.

**Supplementary Information for**

# Dipolar many-body complexes and their interactions in stacked 2D hetero-bilayers

Xueqian Sun[1], Ermin Malic[2] and Yuerui Lu[1,3*]

[1]School of Engineering, College of Engineering and Computer Science, the Australian National University, Canberra, ACT, 2601, Australia

[2]Department of Physics, Philipps-Universität Marburg, 35037 Marburg, Germany

[3]Australian Research Council Centre of Excellence for Quantum Computation and Communication Technology, the Australian National University, Canberra, ACT, 2601 Australia

\* To whom correspondence should be addressed: Yuerui Lu (yuerui.lu@anu.edu.au)

**Supplementary Note 1**

**Background and basic physics**

Excitons, which are bound quasiparticle states composed of electrons and holes, provide an excellent platform for exploring a variety of exotic many-body effects and interacting quantum phenomena, owning to their rich excitonic physics. Excitonic complexes typically dominate and regulate the optical performance of semiconducting materials by stimulating light-matter interactions and coupling. Atomically thin transition-metal dichalcogenides (TMDs) comprise a variety of emerging materials with distinct optical and electronic properties, which have garnered extensive interest as promising contenders for next-generation optoelectronic technologies[1-4], as well as an extraordinary opportunity for condensed matter physics studies[5-8]. In bulk TMDs, the materials are layered with weak van der Waals interactions but exhibit strong bonding within each individual layer, which can be easily isolated into single layers or a few layers with exceptional quality by mechanical exfoliation[2, 9] or electrochemical exfoliation[10]. The 2D feature offers fertile ground for exploring quantum technologies and many-body systems owing to weakened dielectric screening and improved quantum confinement[11-13]. Most monolayer TMDs possess a direct bandgap that falls within the visible to near-infrared spectrum, and the robust excited excitons are formed because of the enhanced attractive Coulomb interactions between electrons and holes, bringing about large exciton binding energies of hundreds of meV[14, 15]. The enormous flexibility of ultra-thin semiconductors, such as their ability to create new physical systems by stacking multiple single layers, is a key advantage over conventional bulk materials. Furthermore, the atomically smooth surface of the TMD layers lacks dangling bonds, which allows the construction of heterostructures by freely stacking them vertically via van der Waals epitaxy without matching their lattices[16]. TMD heterostructures combine the advantages of disparate materials, not only inheriting the already intriguing properties of monolayers, but also enabling a wealth of space to explore new rich physics by expanding the family of double-layer structures[17]. A highly customisable interface is created by van der Waals stacking because it enables the relatively free and loose stacking of individual monolayers, which allows for control over the material type, band alignment, stacking sequence, rotational angle, and application of external fields. Different strategies can be employed to engineer a system by controlling its interfaces. In such heterobilayer structures, the interactions of charge, spin, and lattice structures can generate diverse physical phenomena and multi-particle correlations[18]. The emergence of new excitonic states in TMD heterostructures enables them to hold a significant role in the study of correlations and topology. One novel emergent physical state is the interlayer exciton (IX), in which electrons and holes are allocated to opposite atomic layers. This is formed by the traditional type-II band alignment that most TMD heterobilayers feature[19-22], which accumulates electrons in the lower conduction band minimum (CBM) and collecting holes in the layer with a higher valence band maximum (VBM). This offers a spatial separation between electrons and holes, thus it can also be called indirect excitons. As a result, interlayer excitons host a permeant out-of-plane dipole moment in the vertical direction, which allows for the creation of dipole-dipole interactions[23-25] between IXs. This out-of-plane orientation of the dipoles was experimentally demonstrated via angle-resolved photoluminescence of the $WSe_2/WS_2$ heterostructure. The emission pattern of IX showed a distinct radiation profile that deviated from that of intralayer excitons and resembled that of an in-plane emitter[26]. In particular, the interlayer excitons showed a concave angle-dependent intensity profile, whereas most other excitonic modes demonstrated a convex profile. This facilitates the utilisation of interlayer dipolar excitons in different photonics applications, especially for light-based technologies that rely on light-matter interactions[26]. Because of the nature of spatially separated charges in this stacked structure, other interlayer exciton complexes, including dipolar trions[27, 28] and dipolar biexcitons[29, 30] all carry an electrical dipole moment, which enables their electrical modulation and could significantly affect the exciton system. The concept of a moiré superlattice was introduced to form an in-plane, periodic potential that can trap exciton states, which are closely related to the misalignment in rotation between two stacked monolayers. Some accessible techniques, including high-resolution microscopies and second harmonic generation (SHG), have been widely used to precisely control the rotational angle and positions of stacking layers[31-34]. The unique characteristics of dipolar many-body complexes and the influence of moiré superlattice on them are of paramount importance in a two-dimensional system. Focusing on these aspects opens up new avenues for exploring novel quantum phases and fosters fruitful collaborations between theoretical and experimental researchers.

**Supplementary Note 2**

**Experimental evidence of IX formation and efficiency**

An interlayer exciton (IX) from a TMD heterostructure was first observed and reported in an experiment using $WSe_2/MoSe_2$[35]. Fig. S1a shows the direct observation of IX via photoluminescence (PL) measurements at 20 K. The top and bottom panels show the excitonic spectra from isolated monolayers of $WSe_2$ and $MoSe_2$, and the middle panel illustrates the PL from the heterobilayer, which enables controlled recognition of changes in the spectrum. The distinct feature at 1.40 eV in the spectrum was an emission resulting from interlayer coupling. The intralayer neutral and charged excitons from the monolayers both exist in the hetero-bilayer region with the same emission peak energy as that of the individual monolayers. The insets in the panels display the emission mechanisms from different regions, and the heterostructure preserves both intralayer and interlayer emissions from the excitation area. However, the interlayer exciton emission from the same area at room temperature is faint; it only becomes noticeable and comparable to the intralayer excitons at lower temperatures[35-37]. This is associated with the formation efficiency of interlayer excitons. The IX emission shows a strong twist-angle-dependent performance, reflected by the highest PL intensity occurring at a twist angle near either 0° or 60° and an exponential decrease with a larger twist angle, which shows an almost vanishing intensity when the twist angle exceeds approximately 10°, as displayed in Fig. S1b[38]. The measurement of angle-resolved second-harmonic generation (SHG) is an effective method for determining crystal orientation; thus, this technique has been developed and intensively studied to characterise the rotational alignment of constituent monolayers, thereby identifying the twist angle of hetero-bilayers[31, 39]. This makes the twist angle controllable and offers a perfect platform for designing quantum materials and observing potential physical landscapes. Fig. S1c shows a dramatic enhancement of IX intensities when the temperature decreases; at lower temperatures, one additional IX peak becomes visible and is comparable to the initial peak in the spectra[40]. This phenomenon can be explained by the theoretically predicted thermalisation dynamics of interlayer exciton[41]. The inset summarises the corresponding temperature dependence of the peak position for both emissions, which shows a blue shift as the temperature decreases and follows standard semiconductor behaviour. This could also be good evidence of the uniform and intimate contact between individual monolayers[40]. Besides, Fig. S1d presents a comparison in PL spectrum with different dielectric screening effects to illustrate their impact on the efficiency of IX formation. A strong interlayer exciton and notable intralayer exciton and trion peaks are captured in the photoluminescence spectrum of a non-encapsulated $WSe_2/MoSe_2$ sample, while when the sample is fully encapsulated by hBN dielectric layers, although more spectrally resolved resonance peaks emerge due to the reduced inhomogeneous broadening, the intensity of interlayer excitons is considerably reduced and that of intralayer neutral and charged excitons is improved[42]. The experimental observations of another hetero-bilayer sample with $WS_2/WSe_2$ were consistent with the above illustration. The PL intensity of IX from this sample in the free-standing region at low temperatures was approximately one order of magnitude higher than that of IX from the $SiO_2$-supported area under identical experimental conditions[30]. Fig. S1f is representative evidence to present a more efficient formation of IXs resulting from the lower dielectric screening. At room temperature, the IX emission peaks produced from the type-II band alignment (inset) can be easily observed from a perfectly aligned *R*-type free-standing hetero-bilayer sample, whereas there are no IX peaks from the $SiO_2$-supported region under the same excitation conditions[30].

**Supplementary Note 3**

**Transition in nature of dipolar interaction at small interlayer distance**

Initially, the dipole-dipole repulsive force remains dominant when the dipoles are far apart from each other, leading to an increase in the average system energy as the lateral separation decreases due to the stronger repulsive force. This leads to the unique phenomenon of a blue shift. However, as the dipole separation continues to decrease, the average energy of the system decreases and eventually becomes negative when the dipoles are very close to each other, leading to a transition in the interactions from repulsive to attractive.

**Supplementary Note 4**

**Signatures of moiré excitons**

In addition to direct observation techniques of moiré pattern such as transmission electron microscopy (TEM)[43, 44], scanning transmission electron microscopy (STEM)[45, 46], scanning tunnelling microscopy (STM)[47] and piezoresponse force microscopy (PFM)[48], there is an important signature for moiré excitons in PL measurements that can distinguish them from normal localised excitons. Moiré excitons follow distinct optical selection rules at different positions according to the atomic quantum numbers and relative position between the two layers in real space. In each moiré supercell, the excitons from site $R_h^h$ and $R_h^x$ couple the opposite (σ+ and σ−) polarized light (Fig. S2a), enabling the alternating co- and cross-circularly polarised photoluminescence in the superlattice. In contrast, the site $R_h^M$ is not able to efficiently couple incident light because of its perpendicular dipole in 2D plane[42]. The *H*-typed staking has different selection rules as the triplet states are not considered as dark and have the comparable brightness with spin-singlet ones[49]. Further experimental evidence of moiré-trapped excitons is reflected in the absorption spectrum using reflection contrast spectroscopy. Moiré modulation on the intralayer excitons can be observed as shown in Fig. S2b, multiple prominent resonance peaks are present in the spectrum of excitons if the device is nearly aligned. The evolved multiple states stem from the excitonic transitions that occurred at different moiré induced minibands[50] and were separated over a range of around 100 meV, consistent with the large moiré potential produced in the well-aligned TMD heterostructure. However, these peaks weakened when the twist angle increased and only one peak remained in the large-twist-angle heterobilayer, which was caused by the disappearance of the moiré pattern[31].

**Supplementary Note 5**

**Moiré pattern controlled exciton diffusion**

A similar anomalous diffusion is evident in the 0° sample (Fig. S2c, f), as a completely commensurate structure is formed due to the very close lattice constants of two layers, and the moiré supercell does not exist in this case. As shown in Fig. S2d and e, no obvious diffusion was found beyond the excitation laser point in the sample with a rotational angle of around 1.1° because the IXs were completely localized in the deep moiré potential. However, IX diffusion became observable when the rotational angle changed from 1.1° to 3.5° with the modification of the period of moiré supercell from 20 nm to 5.7 nm. The reduced lateral size of the moiré traps cannot trap as many excitons as before, and the potential depth arising from the superlattice pattern decreases with a larger twist angle, at the same time, the dipole-dipole interactions are enhanced, which allows excitons to tunnel between superlattices, therefore enabling the observation of exciton diffusion.

**Supplementary Note 6**

**Temperature-dependent behaviour of dipolar trions**

From temperature-dependent PL measurements, it can be clearly seen that the PL intensities of the trions decrease significantly with increasing temperature, which results from the thermal dissociation of the trions. This temperature-dependent behaviour is similar to that of intralayer trions in TMD monolayers[51]. The luminescence intensity ratio of trions to excitons for both the direct and indirect states increases with decreasing temperature (Fig. S3a), which follows the mass action law of bound trions[52].

**Supplementary Note 7**

**Additional experimental evidence of high-temperature exciton condensation**

A similar phase diagram with the measured data points acquired from $MoSe_2/WSe_2$ heterobilayer point towards the existence of a transition to a coherent many-body quantum state at around the anticipated critical degeneracy temperature (Fig. S4b). One sharp emission dominated all other optical transitions at cryogenic temperatures, and this exciton state was estimated to exhibit nearly 100% occupation when it entered the degenerate regime, as indicated in Fig. 5c.The full width at half maximum (FWHM) of this emission peak decreased as a function of temperature but reached a minimum value and became nearly constant below the degeneracy temperature (Fig. 5d), although the enhancement was still there for the intensity. The density dependence of the FWHM and peak energy further supports the existence of strong correlated many-body states in this system. The FWHM of the aforementioned sharp emission decreases with increasing exciton density and maintains the lowest value in the degenerate regime; however, it increases dramatically when the density exceeds the critical point (Fig. 5e), and the occupation no longer reaches 100%. This implies the emergence of other optical emissions and the exciton correlations at high densities. A similar transition also occurred at the critical density of the emission energy, which initially shifted slightly and then increased rapidly in the degenerate regime (Fig. 5f), this energy jump can be attributed to the presence of enhanced repulsive dipole-dipole interactions.

**Supplementary Note 8**

**Excitonic optoelectronic devices**

Owing to the numerous excellent characteristics of interlayer excitons discussed above, including tunable emission energy, long lifetime and transport, strong dipolar interactions, and the potential for the realisation of high-temperature Bose-Einstein condensates, interlayer excitons from TMD double layers are a suitable source for use in optoelectronic and excitonic devices. Also because of the 2D nature, IX performance is susceptible to the external environment, which opens opportunities for more potential devices by tailoring the surrounding environment and modifying the integration design. The interactions between excitons and light can be controlled by introducing cavities, as they can change the vacuum field fluctuations and thus modify the exciton-photon coupling[53]. A hetero-bilayer of MoSe$_2$/WSe$_2$ was coupled to a gallium phosphide (GaP) photonic crystal cavity (Fig. S5a) to demonstrate Purcell enhancement[54]. IX-cavity coupling with a considerable enhancement of the photoluminescence (PL) of around 15-fold in intensity (Fig. S5a, bottom) was observed in the weak regime, stemming from an approximately 60-fold Purcell enhancement because IX is in resonance with the photonic crystal cavity. This is favourable for energy-efficient photonic devices because significant light-matter interaction is easy to achieve owing to the small mode volume and high quality factor of the cavities. Also thanks to the cavity, the PL decay curve accompanied by the total intensity of the IXs can be modified by up to two orders of magnitude based on the photonic environment[55] (Fig. S5b), such as the cavity length. Cavities can be involved in interlayer excitons from the type-II band alignment to create lasers. Spatially coherent lasing was first measured at low temperatures from a rotationally well-aligned 2D TMD heterobilayer integrated into a silicon nitride (SiN) grating resonator[56] (Fig. S5c, top panel), forming a direct bandgap of interlayer excitons. A three-level system was provided (Fig. S5c, bottom left) that allows the efficient pumping of intralayer excitons and ultrafast charge transfer; the reduced bandgap and long IX lifetimes facilitate the population inversion for lasing[56]. The oscillator strength of IX is two to three orders in magnitude lower than that of intralayer excitons[57], but the enhancement in the PL yield with a cavity can circumvent the problem of low oscillator strength, improving the coupling sufficiently to further allow population inversion[56, 58] (Fig. S5c, bottom right). The onset of lasing was discerned by observing a super-linear upsurge in the intensity of the emitted radiation at the threshold with a sharp decline in the linewidth (Fig. S5d), which supports an abrupt enhancement in the spatial coherence of the emission around the lasing threshold, indicating that dipolar excitons can act as an efficient gain medium for lasing emission. Furthermore, an infrared room-temperature nanocavity laser was developed based on the IXs from a $MoS_2$/$WSe_2$ heterobilayer on a silicon photonic crystal platform[58] (Fig. S5e). The signature laser properties were clearly observed from the device when the excitation spot moved onto the cavity. The spectrum showed a distinct sharp shape for the emission on the cavity (Fig. S5f), and the emission intensity in the cavity increased faster when the excitation power was above the threshold, whereas no similar indication was observed for spontaneous emission (Fig. S5g). These studies established a good platform for developing coherent light sources for integrated optoelectronics. Another promising van der Waals on-chip integrated photonic device was studied using a waveguide-coupled disk resonator fabricated using hBN as the resonant material[59] (Fig. S5h, inset). The TMD heterobilayer was sandwiched between two hBN slabs and subsequently structured into the photonic circuit. IX can be excited by light near the edge of the disk and coupled out by the waveguide. This design allows the heterostructure to be placed at the position with the strongest optical field intensity, therefore maximising mode overlap and light–matter interactions[59]. This differs from previous studies that transferred the material onto a prefabricated photonic element; the spectrum features a reduced background because it was fully isolated from the free-space PL emissions. Despite the cavity-based integrated devices, photodetectors, acting as essential components of integrated optoelectronics, can also be realised based on interlayer excitons. The low operating temperature required for exciton-based transistors in bulk semiconductor-coupled quantum wells limits their practical application[60-62], and a room temperature IX-based transistor with a long-lived nature was proposed from an encapsulated $MoS_2$/$WSe_2$ heterostructure[63]. An excitonic switch was demonstrated using electrical control (Fig. S5i, top). When the applied fields are absent, IX from the active materials diffuses far away from the excitation point and reaches a recombination site a few micrometres away; however, the light emission could be suppressed by introducing a potential barrier because the motion of IX was impeded (Fig. S5i, bottom), corresponding to the OFF state of the excitonic transistor. The output emissions of the electronic field-effect transistor (FET) can be effectively modulated using this approach. On the basis of interlayer exciton applications, moiré excitons have been further investigated and demonstrated in integrated optoelectronics such as quantum emitters[64] and electronically controlled excitonic switches[65] with polarisation switching. The comprehensive fine-tuning of the emission from moiré excitons in a $WSe_2/MoSe_2$ heterobilayer was recently revealed, and the emission of quantum light was observed from the photon anti-bunching[64]. Another excitonic switch very similar to the aforementioned excitonic phototransistor was realised with moiré excitons at low temperatures[65]. A polarisation switch with an adjustable emission intensity and energy was enabled by electrically controlled tuning. Interestingly, the exciton density can be enhanced by an order of magnitude when an electrostatic trap is applied to the valley-polarised IX[66], which offers more possibilities for the realisation of the coherent quantum state of valley-polarised excitons through Bose–Einstein condensation. These studies have assisted in understanding the quantum nature of moiré emitters and demonstrated promising progress towards the achievement of macroscopic quantum states of moiré excitons.

**Supplementary Note 9**

**Motivation of the study**

The long range, fast transport and tunable energy of coherent excitons are of paramount significance for advancing quantum computing applications, signal processing devices, and high-speed excitonic circuits, etc. These are coincidentally the basic and unique characteristics of interlayer dipolar excitons. They are recently known to undergo Bose-Einstein condensation and have the capacity of forming the quantum liquids that show quantum mechanical effects at the macroscopic level, such as ultracold atoms, superconductors, and superfluids. The notion of superconductivity or superfluidity refers to the physical characteristics in which the electrical resistance vanishes and the viscosity of a fluid is zero; therefore, the fluid travels without any energy loss. They are normally realised below a critical temperature and superconductors also expel magnetic fields when the materials reach the superconducting state[67]. In recent decades, pushing the critical temperature to a higher level has by optimising the controlling parameters or tailoring the system design has been a hot topic in quantum research. This eliminates the requirements for special equipment or environments for cooling and promotes the implementation of practical applications. If the aforementioned quantum phenomena can be utilised commercially, a new generation of novel high technology will be developed, enabling many new possibilities for highly efficient electronics with lower costs and emissions, such as, ultrafast and energy-efficient computer chips, highly efficient electricity grids, low-latency broadband wireless communications, ultra-powerful magnets, highly efficient magnetic levitating trains, and even electric airplanes. It is also beneficial for biomedical and security applications, because it can enable the development of high-resolution imaging techniques and emerging sensors. This will be an industrial revolution and have a major impact on contemporary optoelectronic devices.

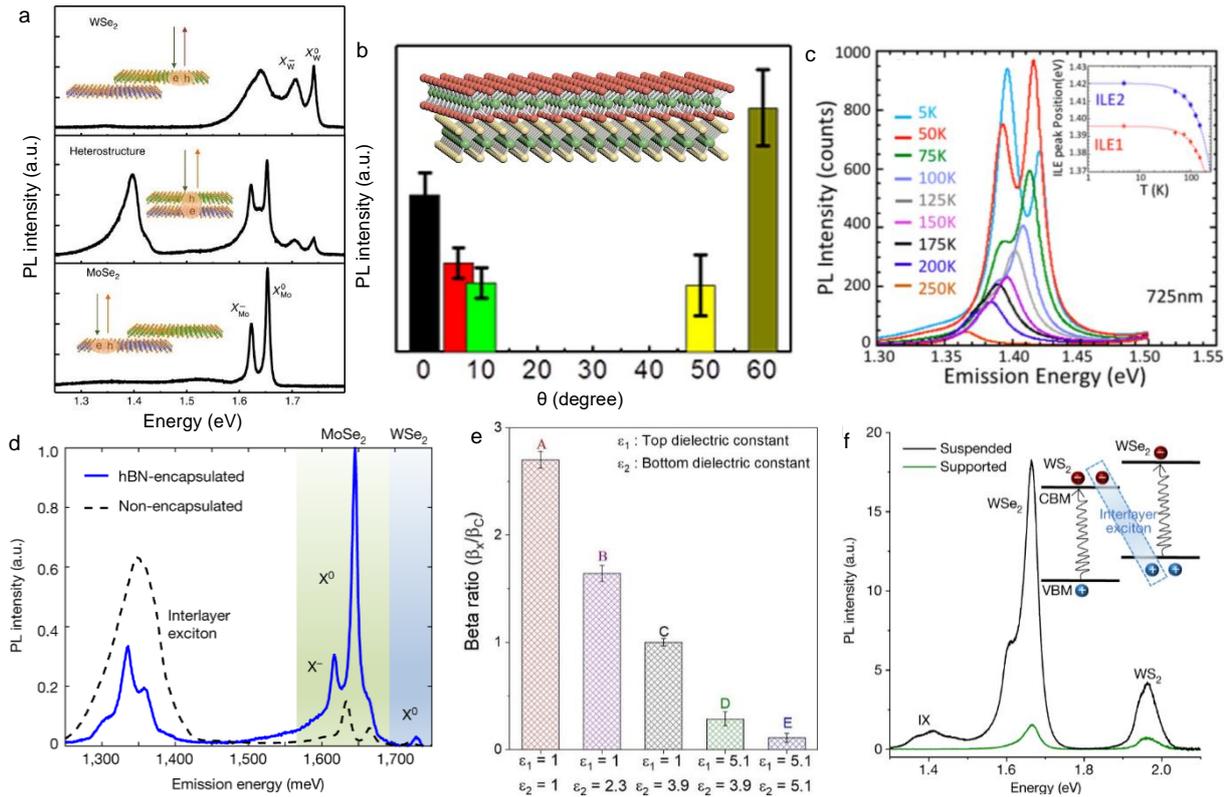

**Supplementary Fig S1. │ Interlayer exciton formation efficiency. a,** Measured photoluminescence (PL) of individual monolayers and the constructed heterstrucure from WSe$_2$/MoSe$_2$ heterobilayer at 20 K, illustrating the emergence of interlayer exciton. **b,** Twist-angle dependent PL intensity of the interlayer exciton formed from MoSe$_2$/WSe$_2$ heterobilayers. **c,** Temperature dependent PL spectra of interlayer exciton from MoSe$_2$/WSe$_2$ heterostructure. The inset shows the peak shift for each emission line under different temperatures. **d**, Comparison of the PL spectrum from an encapsulated (solid blue curve) and non-encapsulated (dashed black curve) MoSe$_2$/WSe$_2$ heterostructure. **e**, Formation efficiency (represented by the extracted beta ratio) of interlayer exciton as a function of the top and bottom dielectric constants. **f**, PL spectra from the SiO$_2$-supported and suspended WS$_2$/WSe$_2$ hetero-bilayers at room temperature. IX emissions were observed from suspended region with the reduced dielectric screening, but no emission was observed from the SiO$_2$-supported sample. The inset is the schematic diagram displaying the formation of interlayer exciton between TMD monolayers. Reproduced with permission from: **a**, ref.[35], Springer Nature Ltd; **c**, ref.[40], American Chemical Society; **d**, ref.[42], Springer Nature Ltd; **e**, **f**, ref.[30], Springer Nature Ltd. Adapted with permission from: **b**, ref.[38], American Chemical Society.

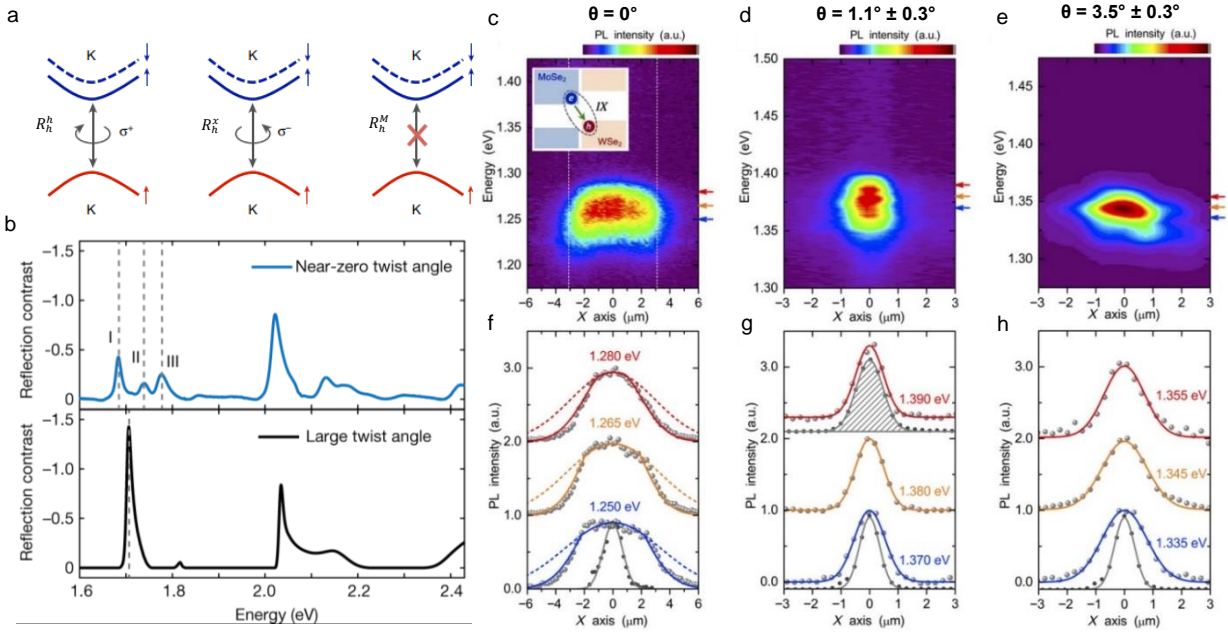

**Supplementary Fig S2.** | **Moiré exciton and moiré impeded diffusion. a**, Interlayer exciton optical selection rules for small twisted MoSe$_2$/WSe$_2$ hetero-bilayer in K-valley at atomic registries. Exciton emission at $R_h^h$ ($R_h^x$) is left-circularly (right-circularly) polarized. Emission from the $R_h^M$ site is dipole-forbidden under normal incidence. **b**, Moiré modulation on intralayer excitons indicated in the reflectance measurement. **c**, Spatially resolved PL image of interlayer excitons in MoSe$_2$/WSe$_2$ bilayers fabricated by CVD growth with zero twist rotation. **d**, **e**, Spatially resolved PL images of interlayer excitons in MoSe$_2$/WSe$_2$ bilayers fabricated by mechanical exfoliation and dry transfer methoed with twist angle of $1.1° \pm 0.3°$ and $3.5° \pm 0.3°$. There is no diffusion beyond the excitation laser point in (**d**) and IX diffusion beyond the laser spot size becomes observable in (**e**). **f-h**, Extracted corresponding PL spectra from (**c-e**) at color indicted energies. The PL line profiles in (**f**) are truncated by the boundaries of the hetero-bilayer region (the white dashed lines in **c**). The 660 nm excitation laser profile and 900 nm laser profile are indicated with grey solid lines at bottom and top grey shaded area in (**g**), respectively. The size of the IX (IXs near 900 nm) PL spot appears slightly larger than the excitation laser (at 660 nm) spot size because of the imaging optics at different wavelengths. Measured data points are fitted by a Gaussian function. Reproduced with permission from: **b**, ref.[31], Springer Nature Ltd; **c-h**, ref.[68], The American Association for the Advancement of Science. Adapted with permission from: **a**, ref.[42], Springer Nature Ltd.

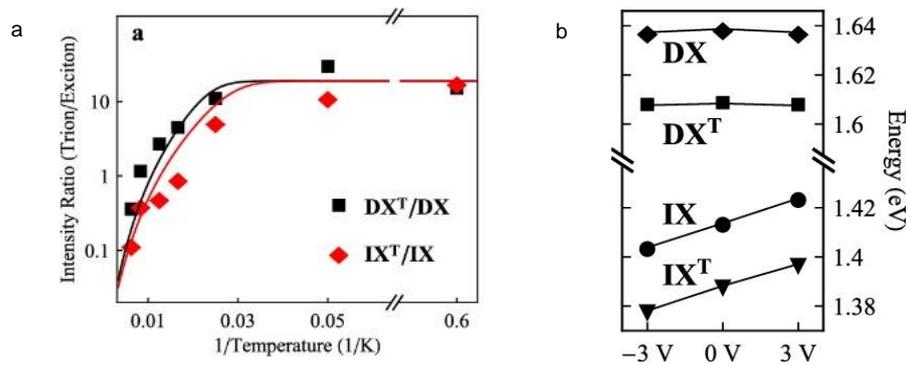

**Supplementary Fig S3.│ Temperature- and gate-dependent behaviour of direct and indirect excitons and trions. a**, Extracted experimental (symbols) and simulated (lines) integrated luminescence intensity ratio (Trion/Exciton) of both interlayer ($IX^T$/ IX, red) and intralayer ($DX^T$/ DX, black) emissions as a function of 1/T. **b**, Gate-dependent peak energy of interlayer ($IX^T$, IX) and intralayer ($DX^T$, DX) PL emissions. Reproduced with permission from: **a-b**, ref.[28], American Chemical Society.

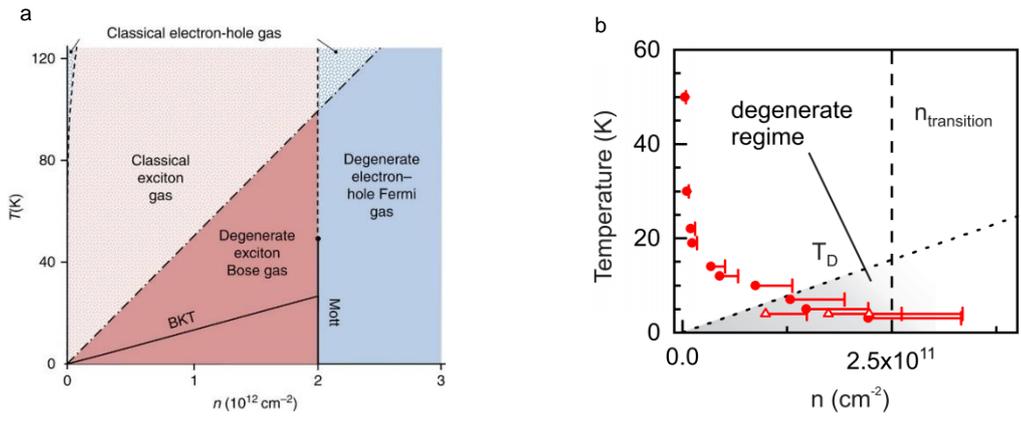

**Supplementary Fig S4.│ Exciton Condensation in stacked 2D bilayers. a**, Phase diagram of excitons with the variation of temperature and exciton densities. **b**, Many-body phase diagram with measured data from MoSe$_2$/WSe$_2$ heterobilayer, showing the quantum degeneracy of IX at reduced temperatures and sufficiently high exciton densities. Reproduced with permission from: **a**, ref.[5], Springer Nature Ltd; **b**, ref.[69], American Physical Society.

a
b
c
d
e
f
g
h
i

**Supplementary Fig S5.│Device integration and applications. a**, Side view of a device by palcing MoSe$_2$/WSe$_2$ heterobilayer on top of a GaP photonic crystal (top). Photoluminescence (PL) spectra from the device on and off the defect cavity in the photonic crystal (bottom). **b**, Cavity controlled lifetime and emission width of IXs with different cavity lengths. **c**, Room temperature lasing by integrating IXs from MoSe$_2$/WSe$_2$ heterostructure with grating cavities (top). The heterobilayer form a type-II band alignment, which leads to a three-level system for lasing (bottom left). Interlayer exciton lasing emission from the laser device (bottom right). **d**, Excitation power dependent photon occupancy (red) and linewidth (blue) of the transverse electric (TE) emission from the device, displaying lasing characteristics. **e**, Fabricated nanolaser from MoS$_2$/WSe$_2$ heterobilayer on a photonic crystal cavity. **f**, Room-temperature interlayer exciton lasing from the device shown in **e**, the inset shows the spontaneous emission of the interlayer exciton outside the cavity region. **g**, PL output intensities of the device as a function of the excitation pump powers. The red curve shows the cavity interlayer exciton emission with a kink, stating the onset of lasing operation. The black curve is the spontaneous emission, which shows a linear dependence on the pump power. **h**, Cavity-coupled photoluminescence spectrum showing distinct emissions separated by the free-spectral range of the whispering gallery modes (WGMs) of the disk resonator. The insets illustrate the device made of a waveguide-coupled disk resonator with an integrated MoSe$_2$/WSe$_2$ hetero-bilayer. The interlayer exciton emission is coupled to whispering gallery modes (WGMs) of the disk resonator, which is excited from one end and detected at the other end of the waveguide. **i**, Schematic illustration of the excitonic switch made from hBN encapsulated MoS$_2$/WSe$_2$ heterostructure (top). The three control gates are applied to engineer the transport of interlayer excitons through the device. Photoluminescence images of the interlayer exciton emission in ON and OFF states (bottom). Dashed lines outline the areas of samples and the red circle is the excitation spot. Reproduced with permission from: **a**, ref. [54], IOP Publishing; **b**, ref. [55], Springer Nature Ltd; **c**, **d**, ref. [56], Springer Nature Ltd; **e-g**, ref. [58], The American Association for the Advancement of Science; **i**, ref. [63], Springer Nature Ltd. Adapted with permission from: **h**, ref. [59], American Chemical Society.